\documentclass[a4paper,12pt]{article}

\pdfoutput=1
\usepackage{amsmath}
\usepackage{amssymb}
\usepackage{amsfonts}
\usepackage{mathrsfs}
\usepackage{bbm}
\usepackage{graphicx,subfigure}

\usepackage{bookmark}

\usepackage{cite}

\usepackage{calc}
\usepackage{hyperref}
\usepackage{array}

\usepackage{ulem}

\usepackage{multirow}
\usepackage{color}
\usepackage{xcolor,colortbl}

\usepackage{psfrag}
\usepackage{pstricks}
\usepackage{epsfig}
\numberwithin{equation}{section}
\allowdisplaybreaks

\newlength{\dinwidth}
\newlength{\dinmargin}
\definecolor{nicered}{rgb}{1.0,0.0,0.2}

\definecolor{color1}{rgb}{0.9,.4,.2}

\definecolor{color2}{rgb}{0.3,.6,.7}

\definecolor{color3}{rgb}{0.7,.2,.7}

\usepackage{color}

\begin{document}

\title{
\vspace*{-0.5cm}
\bf \Large
Study of  $\mathbf{\boldsymbol{D \rightarrow a_0 (980) e^+ \nu_e}}$  decay in the light-cone sum rules approach}

\author{Xiao-Dong Cheng$^{1}$\footnote{chengxd@mails.ccnu.edu.cn}, Hai-Bo Li$^{2,3}$\footnote{lihb@ihep.ac.cn}, Bin Wei$^{1}$\footnote{weibinwuli@163.com}, Yu-Guo Xu$^{1}$\footnote{yuanguoxv@163.com}, Mao-Zhi Yang$^{4}$\footnote{yangmz@nankai.edu.cn}\\
\\
{$^1$\small College of Physics and Electronic Engineering,}\\[-0.2cm]
{    \small Xinyang Normal University, Xinyang 464000, People's Republic of China}\\[-0.1cm]
{$^2$\small  Institute of High Energy Physics,}\\[-0.2cm]
{    \small  Beijing 100049, People's Republic of China}\\[-0.1cm]
{$^3$\small  University of Chinese Academy of Sciences,}\\[-0.2cm]
{    \small  Beijing 100049, People's Republic of China}\\[-0.1cm]
{$^4$\small School of Physics,}\\[-0.2cm]
{    \small Nankai University, Tianjin 300071, People's Republic of China}\\[-0.1cm]}

\date{}
\maketitle
\bigskip\bigskip
\maketitle
\vspace{-1.2cm}

\begin{abstract}
{\noindent}Within the QCD light-cone sum rule (LCSR) approach, we investigate the transition form factors of $D\rightarrow a_0(980)$ up to the twist-3 light-cone distribution amplitudes (LCDAs) of the scalar meson $a_0(980)$ in the two-quark picture. Using these form factors, we calculate the differential decay widths and branching ratios of the $D\rightarrow a_0(980) e^+ \nu_e$ semileptonic decays. We obtain ${\mathcal B}(D^0\rightarrow a_0^- (980) e^+ \nu_e)=(4.08^{+1.37}_{-1.22})\times 10^{-4}$ and ${\mathcal B}(D^+\rightarrow a_0^0 (980) e^+ \nu_e)=(5.40^{+1.78}_{-1.59})\times 10^{-4}$. The results are sensitive to the $a_0(980)$ inner structure.  These decays can be searched for at BESIII experiment, and any experimental observations will be useful to identify internal quark contents of the $a_0(980)$ meson, which will shed light on understanding theoretical models.
\end{abstract}

\newpage

\section{Introduction}
\label{sec:intro}
The property of the light scalar meson $a_0(980)$ has been controversial for over three decades, which is one of the alluring issues in light hadron spectroscopy. Currently two scenarios are suggested~\cite{Cheng:2005nb,Sun:2010nv}. In scenario 1, $a_0(980)$ is treated as the lowest lying $q\bar{q}$ states, and $a_0(1450)$ as the corresponding first excited state. In scenario 2, $a_0(1450)$ is assumed to be the lowest lying $q\bar{q}$ resonances and the corresponding first excited state lies between $(2.0-2.3)$ GeV, while $a_0(980)$ is taken to be the member of a four-quark nonet. Due to the absence of convincing evidence both experimentally and theoretically~\cite{Mathur:2006bs,Prelovsek:2010kg,Alexandrou:2012rm,Cheng:2013fba}, the nature of the isovector states $(a_0^+(980),a_0^0(980),a_0^-(980))$ is still in ambiguity.

The $B$ decays involving $a_0 (980)$ have been studied extensively~\cite{Cheng:2005nb,Sun:2010nv,Cheng:2013fba,Cheng:2003sm,Wang:2008da}, but it is still difficult to draw a conclusion whether $a_0 (980)$ is a 2-quark or a 4-quark state. In order to understand the $a_0(980))$ structure,
more decays involving $a_0 (980)$ in experiment and more investigations in theoretical methods are needed.
In this paper we study the $a_0(980)$ production in $D \to a_0(980)$ semileptonic decays.  The main difficulty is to properly evaluate the hadronic matrix elements for $D \rightarrow a_0 (980)$ transition. The  form factors are generally governed by non-perturbative QCD dynamics. There are several methods to deal with the difficulty, such as the quark model~\cite{Wirbel:1985ji}, the light-front approach~\cite{Choi:1999nu,Zhang:1994hg,Cheung:1995ub}, QCD sum rule (QCDSR)~\cite{Shifman:1978by,Novikov:1981xi}, light-cone QCD sum rule~(LCSR)~\cite{Balitsky:1989ry,Braun:1988qv,Chernyak:1990ag}, perturbative QCD factorization approach~\cite{Keum:2000ph,Keum:2000wi,Lu:2000em}. The LCSR approach, which starts with the operator product expansion~(OPE) of a two-point correlation function near the light cone $x^2 = 0$ and with the help of the hadronic dispersion relation and quark-hadron duality, is successfully used to calculate heavy-to-light form factors in the region of small momentum transfer squared, $q^2 = (p_{P} - p_{a_0(980)})^2$. In the LCSR approach, the sum rules for the form factors are functions of the light-cone distribution amplitudes~(LCDAs) of the scalar meson, which can be expanded into a series of Gegenbaur polynomials. At present, both the twist-2 and twist-3 LCDAs of  the scalar mesons have been investigated~\cite{Cheng:2005nb,Lu:2006fr,Han:2013zg} based on the QCD sum rules.
Here we would like to adopt LCSR approach to study the semileptonic decays $D^0\rightarrow a_0^- (980) e^+ \nu_e$ and $D^+ \rightarrow a_0^0 (980) e^+ \nu_e$, which can give a hint on the inner structure of $a_0 (980)$. We will calculate the branching ratio of these decays under the assumption that $a_0(980)$ is the lowest lying $q\bar{q}$ states.

From experimental  side, the CLEO-c experiment at the Cornell Electron Storage Ring (CESR) $e^+e^-$ collider collected a sample of $5.31\times 10^6$ $D\bar{D}$ pairs~\cite{Onyisi:2013bjt,Bonvicini:2013vxi}, and the BESIII experiment at BEPCII has also accumulated a sample of $19.4 \times 10^6$ $D\bar{D}$ pairs near the $D\bar{D}$ threshold~\cite{Asner:2008nq,Cheng:2007uj,Ablikim:2015wel}.   These data samples provide an ideal place to study the $D\rightarrow a_0 (980)$ semileptonic decays and investigate the nature of the isovector states $a_0(980)$.

The paper is organized as follows. In Section 2, we briefly introduce the flavor wave functions of $a_0 (980)$ and the Gengenauber moments of twist-2 and twist-3 distribution amplitudes in the QCD sum rules. In section 3, we present the effective Hamiltonian responsible for $c\rightarrow d$ transition in the Standard Model and the parameterizations of hadronic matrix elements, then the sum rules for the form factors on the light-cone are presented with the standard correlation function to the leading Fork state. In section 4, the numerical computations of form factors are performed with the input parameters. Subsequently, we analyze the differential decay rates and the branching ratios of $D^0\rightarrow a_0^- (980) e^+ \nu_e$ and $D^+ \rightarrow a_0^0 (980) e^+ \nu_e$ by applying the form factors. The last section is reserved for the conclusion. The explicit expressions of the Gengenauber moments of twist-3 distribution amplitudes are collected in the Appendix.

\section{Physical properties of $a_0 (980)$ }
\label{sec:properties}
The scalar $a_0 (980)$ is an isovector state, and its structure is still not well established. There are two possible scenarios for the quark content of $a_0 (980)$, which have been stated in Sec.~\ref{sec:intro}. In the four-quark scenario, the flavor wave functions of $a_0 (980)$ read~\cite{Cheng:2005nb,Wang:2009azc}
\begin{align}\label{eq:fourflavorfunc}
| a_0^0 (980)\rangle =\frac{1}{\sqrt{2}}| (u\bar{u} -d\bar{d})s\bar{s}\rangle\,, \qquad | a_0^- (980)\rangle =| d\bar{u}s\bar{s}\rangle\,,\qquad| a_0^+ (980)\rangle =| u\bar{d}s\bar{s}\rangle.
\end{align}
In the $q\bar{q}$ picture, $a_0 (980)$ is viewed as P-wave state and its flavor wave functions are given by
\begin{align}\label{eq:twoflavorfunc}
| a_0^0 (980)\rangle =\frac{1}{\sqrt{2}}(| u\bar{u}\rangle-| d\bar{d}\rangle)\,, \qquad | a_0^- (980)\rangle =| d\bar{u}\rangle\,,\qquad| a_0^+ (980)\rangle =| u\bar{d}\rangle.
\end{align}
Up to the leading Fock states, the light-cone distributions of $a_0 (980)$ made up of $q_2\bar{q}_1$ can be defined as
\begin{align}
&\left\langle a_0 (980)\left( p\right) \left| \bar{q}_{2}\left( x\right)
\gamma _{\mu }q_{1}\left( y\right) \right| 0\right\rangle
=p_{u}\int_{0}^{1}due^{i\left( up\cdot x+\bar{u}p.y\right) }\Phi
_{S}\left( u,\mu \right),  \nonumber \\
&\left\langle a_0 (980)\left( p\right) \left| \bar{q}_{2}\left( x\right)
q_{1}\left( y\right) \right| 0\right\rangle
=m_{a_0 (980)}\int_{0}^{1}due^{i\left( up\cdot x+\bar{u}p.y\right) }\Phi
_{S}^{s}\left( u,\mu \right),  \label{DAs}\\
&\left\langle a_0 (980)\left( p\right) \left| \bar{q}_{2}\left( x\right)
\sigma _{\mu \nu }q_{1}\left( y\right) \right| 0\right\rangle
=-m_{a_0 (980)}\left(
p_{\mu }z_{\nu }-p_{\nu }z_{\mu }\right) \int_{0}^{1}due^{i\left( up\cdot x+%
\bar{u}p.y\right) }\Phi _{S}^{\sigma }\left( u,\mu \right),
\nonumber
\end{align}
where $q_2\bar{q}_1$ denotes the quark content for $a_0^0 (980)$, $a_0^+ (980)$ and $a_0^- (980)$, which should be $d\bar{d}$, $u\bar{d}$ and $d\bar{u}$ respectively. Note that the factor $\frac{1}{\sqrt{2}}$ involved in the flavor wave function of $a_0^0 (980)$ in Eq.~(\ref{eq:twoflavorfunc}) have not been taken into account here. It should be included in the calculation of the final decay rates for $D\rightarrow a_0 (980)$ semileptonic decays. $m_{a_0 (980)}$ is the mass of $a_0 (980)$, $z=x-y$ is the displacement between the quark and anti-quark, and $u$ is the momentum fraction carried by the quark $q_2$ in $a_0 (980)$ with $\bar{u}=1-u$. $\Phi _{S}\left( u,\mu \right)$ and $(\Phi _{S}^{s}\left( u,\mu \right),\Phi_{S}^{\sigma }\left( u,\mu \right))$ are the twist-2 and twist-3 distribution functions, which can be expanded into a series of Gengenbauer polynomials in the Hilbert
space~\cite{Braun:2003rp,Chernyak:1983ej}
\begin{align}
&\Phi _{S}\left( u,\mu \right) =\bar{f}_{S}\left( \mu \right) 6u\bar{u}%
\left[ B_{0}\left( \mu \right) +\sum_{m=1}^{\infty }B_{m}\left( \mu
\right)
C_{m}^{3/2}\left( 2u-1\right) \right],  \nonumber \\
&\Phi _{S}^{s}\left( u,\mu \right) =\bar{f}_{S}\left( \mu \right) \left[
1+\sum_{m=1}^{\infty }a_{m}\left( \mu \right) C_{m}^{1/2}\left( 2u-1\right) %
\right] , \label{Gegexpansion} \\
&\Phi _{S}^{\sigma }\left( u,\mu \right)=\bar{f}_{S}\left( \mu \right) 6u%
\bar{u}\left[ 1+\sum_{m=1}^{\infty }b_{m}\left( \mu \right)
C_{m}^{3/2}\left( 2u-1\right) \right],  \nonumber
\end{align}
where $\bar{f}_{S}$ is the scalar decay constant, which is determined by
\begin{align}
\left\langle a_0 (980)\left( p\right) \left| \bar{q}_{2}q_{1}\right|
0\right\rangle =m_{a_0 (980)}\bar{f}_{S},
\end{align}
$C_{m}^{3/2,1/2}\left( 2u-1\right)$ are Gengenbauer polynomials.
$B_m$, $a_m$ and $b_m$ are the Gengenbauer moments for twist-2 and twist-3 LCDAs respectively. With the orthogonality of Gengenbauer polynomials
\begin{align}
&\int^1_0  du C_n^{1/2}(2u-1)C_m^{1/2}(2u-1)=\frac{1}{2 n+1}\delta_{m n}, \nonumber \\
&\int^1_0  du u(1-u)C_n^{3/2}(2u-1)C_m^{3/2}(2u-1)=\frac{(n+2)(n+1)}{4 (2n+3)}\delta_{m n} \nonumber ,
\end{align}
the first four Gengenbauer moments read
\begin{align}
& a_1\left( \mu \right)=3 \langle \xi_s^1\rangle, \; a_2\left( \mu \right)=\frac{5}{ 2} \left( 3 \langle \xi_s^2\rangle-1 \right), \nonumber \\
&a_3\left( \mu \right)=\frac{7}{ 2} \left( 5 \langle \xi_s^3\rangle-3\langle \xi_s^1\rangle \right), \;
 a_4\left( \mu \right)=\frac{9}{ 8} \left( 35 \langle \xi_s^4 \rangle-30 \langle \xi_s^2 \rangle+3 \right),\label{relationA}\\
& b_1\left( \mu \right)=\frac{5}{3} \langle \xi_{\sigma}^1 \rangle, \; b_2\left( \mu \right)=\frac{7 }{ 12} \left( 5\langle \xi_{\sigma}^2\rangle -1 \right), \nonumber\\
& b_3\left( \mu \right)=\frac{3}{ 4} \left( 7 \langle \xi_{\sigma}^3\rangle-3\langle \xi_{\sigma}^1\rangle \right), \;
b_4\left( \mu \right)=\frac{11}{ 24} \left(21 \langle \xi_{\sigma}^4\rangle -14 \langle \xi_{\sigma}^2 \rangle+1 \right) ,\label{relationB}\\
& B_0\left( \mu \right)= \langle \xi_{\phi}^0\rangle, \;B_1\left( \mu \right)=\frac{5}{3} \langle \xi_{\phi}^1 \rangle, \; B_2\left( \mu \right)=\frac{7 }{ 12} \left( 5\langle \xi_{\phi}^2\rangle -\langle \xi_{\phi}^0\rangle \right), \nonumber\\
& B_3\left( \mu \right)=\frac{3}{ 4} \left( 7 \langle \xi_{\phi}^3\rangle-3\langle \xi_{\phi}^1\rangle \right), \;
B_4\left( \mu \right)=\frac{11}{ 24} \left(21 \langle \xi_{\phi}^4\rangle -14 \langle \xi_{\phi}^2 \rangle+\langle \xi_{\phi}^0\rangle \right),\label{relationC}
\end{align}
where $\langle \xi_{\phi}^0\rangle=\left({m_2 (\mu)-m_1 (\mu)}\right)/{m_{a_0 (980)}}$ is the zeroth moment of twist-2 LCDAs with $m_{1}$ and $m_{2}$ being the masses of quarks $q_{1}$ and $q_{2}$, respectively. Moreover, various moments $\langle \xi_{\phi}^n\rangle$, $\langle \xi_s^n\rangle$ and $\langle \xi_{\sigma}^n\rangle$ for both twist-2 and twist-3 LCDAs have been computed in Refs.~\cite{Cheng:2005nb,Lu:2006fr,Han:2013zg} based on QCD sum rules approach. The explicit expressions of $\langle \xi_s^n\rangle$ and $\langle \xi_{\sigma}^n\rangle$ are collected in the Appendix.
\section{The light-cone sum rule for $D\rightarrow a_0 (980)$ transition factors}
In the standard model (SM), the effective Hamiltonian for the $c\rightarrow d e^+ \nu_e$ transition is
\begin{align}
 \mathcal{H}_{eff}(c \to d e^+ \nu_e)=\frac{G_{F}}{\sqrt{2}}V_{cd}^{*}
 \bar{d}\gamma_{\mu}(1-\gamma_5)c \,
 {\bar{\nu}}_e\gamma^{\mu}(1-\gamma_5)e^+ +h.c. \, ,
 \label{effectiveH b to u}
\end{align}
where $V_{cd}$ is the Cabibbo-Kobayashi-Maskawa (CKM) matrix element. In order to calculate the decay amplitude for the semi-leptonic decay of $D\rightarrow a_0(980)$ at hadronic level, the hadronic matrix element $\langle a_0(980)(p)|\bar{d}\gamma_{\mu}\gamma_5 c|D(p+q)\rangle$ need to be evaluated. Because of parity conservation in strong interaction, the vector current does not contribute. The above matrix element can be parameterized in terms of the form factors $f_+ (q^2)$ and $f_- (q^2)$ as
\begin{align}
\langle a_0(980)(p)|\bar{d}\gamma _{\mu }\gamma
_{5}c|D(p+q)\rangle
=-i[f_{+}(q^{2})p_{\mu }+f_{-}(q^{2})q_{\mu }],
\label{axial form factor}
\end{align}
The light-cone sum rules for the form factors $f_+ (q^2)$ and $f_- (q^2)$ can be obtained by introducing a proper chiral correlator. More explicitly, we adopt the following correlator
\begin{align}
\Pi _{\mu }\left( p,q\right) =-\int d^{4}xe^{iqx}\left\langle
a_0(980)\left( p\right) \left| T\left\{ \bar{q}_{2}\left( x\right) \gamma
_{\mu }\gamma _{5} c\left( x\right),\bar{c}\left( 0\right) i\gamma _{5}q_1\left(
0\right)\right\}
\right| 0\right\rangle . \label{correlator1}
\end{align}
With the standard procedure to deal with the correlators which had been used in that of the $B\to$ scalar transition form factors~\cite{Wang:2008da,Balitsky:1987bk,Diehl:1998kh,Khodjamirian:1998ji,Colangelo:2010bg,Hambrock:2015aor}, we can arrive at the sum rules for the $D\rightarrow a_0 (980)$ transition form factor. The light-cone sum rules for the form factors $f_{\pm}(q^2)$ read
\begin{align}
&f_{+}\left( q^{2}\right) =\frac{\left( m_{c}+m_{q_1}\right) }{
m_{D}^{2}f_{D}}\exp \left( \frac{m_{D}^{2}}{M^2} \right)
\bigg \{ \int_{u_{0}}^{1}\frac{du}{u}\exp \left[
-\frac{m_{c}^{2}+u\bar{u}m_{a_0 (980)}^{2}-\bar{u}q^{2}}{uM^2}\right] \times
\bigg[ -m_{c}\Phi _{S}\left( u\right)\nonumber \\
&\hspace{1.5 cm}+ m_{a_0 (980)}\left( u\Phi_{S}^{s}\left( u\right) +\frac{1}{3}\Phi
_{S}^{\sigma }\left( u\right) \right) +\frac{1}{uM^2}\frac{m_{a_0 (980)}}{6}\Phi _{S}^{\sigma
}\left( u\right) \left( m_{c}^{2}-u^{2}m_{a_0 (980)}^{2}+q^{2}\right) \bigg]\nonumber \\
&\hspace{1.5 cm}+\frac{m_{a_0 (980)}}{6}\Phi _{S}^{\sigma }\left( u_{0}\right) \exp \left( -\frac{%
s_{0}}{M^2}\right) \frac{m_{c}^{2}-u_{0}^{2}m_{a_0 (980)}^{2}+q^{2}}{
m_{c}^{2}+u_{0}^{2}  m_{a_0 (980)}^{2}-q^{2}} \bigg \}, \label{fplus}\\
&f_{-}\left( q^{2}\right) =\frac{\left( m_{c}+m_{q_1}\right) }{%
m_{D}^{2}f_{D}}\exp \left( \frac{m_{D}^{2}}{M^2} \right)
\bigg \{ \int_{u_{0}}^{1}\frac{du}{u}\exp \left[
-\frac{m_{c}^{2}+u\bar{u} m_{a_0 (980)}^{2}-\bar{u}q^{2}}{uM^2}\right]  \nonumber \\
&\hspace{1.5 cm}\times
\bigg[ \bigg( m_{a_0 (980)}\left( \Phi _{S}^{s}\left( u\right)
+\frac{1}{6u}\Phi _{S}^{\sigma }\left(
u\right) \right) \bigg) -\frac{1}{u^{2}M^2}\frac{m_{a_0 (980)}}{6}\Phi _{S}^{\sigma
}\left( u\right) \left( m_{c}^{2}+u^{2}m_{a_0 (980)}^{2}-q^{2}\right)\bigg]\nonumber \\
&\hspace{1.5 cm}-\frac{m_{a_0 (980)}}{6u_{0}}\Phi _{S}^{\sigma }\left( u_{0}\right) \exp
\left(-\frac{s_{0}}{M^2}\right)\bigg \}, \label{fminus}
\end{align}
and
\begin{align}
u_{0}={\frac{-(s_{0}-q^{2}-m_{a_0 (980)}^{2})+\sqrt{%
(s_{0}-q^{2}-m_{a_0 (980)}^{2})^{2}+4m_{a_0 (980)}^{2}(m_{c}^{2}-q^{2})}}{2m_{a_0 (980)}^{2}},}
\label{unot}
\end{align}
where $c{\bar{q}}_1$ denotes the quark content for $D^0$ and $D^+$, which are $c\bar{u}$ and $c\bar{d}$, respectively. $s_{0}$ is the threshold parameter corresponding to D channel. From the sum rules in Eqs.(\ref{fplus}) and
(\ref{fminus}), we can see that the form factor $f_{+}(q^2)$ receives contributions from both twist-2 and twist-3 distribution amplitudes of $a_0 (980)$, while only twist-3 LCDA contribute to the form factor $f_{-}(q^2)$. Considering Eqs. (\ref{effectiveH b to u}) and (\ref{axial form factor}), we can obtain the differential decay rate
for the $D^0\to a_0^- (980) e^+ \nu_e$, which is expressed as
\begin{align}
\frac{d\Gamma}{dq^2}(D^0\to &a_0^- (980) e^+ \nu_e)=\frac{G_F^2|V_{cd}|^2}{768 \pi^3
 m_D^3}\frac{(q^2-m_e^2)^2}{q^6}\sqrt{(m_D^2+m_{a_0 (980)}^{2}-q^2)^2-4m_D^2m_{a_0 (980)}^{2}}\nonumber\\
&\times\bigg \{\left(f_{+}(q^2)\right)^2\bigg[(q^2+m_{a_0 (980)}^{2}-m_D^2)^2 (q^2+2m_e^2)-q^2m_{a_0 (980)}^{2}(4q^2+2m_e^2)\bigg]\nonumber\\
&+6f_{+}(q^2)f_{-}(q^2)q^2 m_e^2(m_D^2-m_{a_0 (980)}^{2}-q^2)+6\left(f_{-}(q^2)\right)^2q^4 m_e^2\bigg \},\label{eq:widthsemi}
\end{align}
where $m_e$ is the mass of electron and the effective region of $q^2$ is $m_e^2 \leq q^2\leq (m_D-m_{a_0 (980)})^2$. As for the differential decay rates for the $D^+\to a_0^0 (980) e^+ \nu_e$, a factor $1/2$ should be included because of the quark content of $a_0^0 (980)$ in Eq. (\ref{eq:twoflavorfunc}).
\section{Numerical calculation and discussion}
For the numerical analysis, the following input parameters are collected \cite{Narison:2010wb,Narison:2011rn,Olive:2016xmw}
\begin{align}
\begin{array}{ll}
G_F=  1.1663787\times 10^{-5} {\rm{GeV}^{-2}}, &  |V_{cd}|=0.219\pm{0.006},\\
m_c=(1.275 \pm 0.025) {\rm{GeV}}, & m_u=(2.5\pm 0.8) \ {\rm{MeV}},\\
m_d=(5.0\pm 0.8) \ {\rm{MeV}}, &  m_{D}=(1.86484\pm0.00005) {\rm{GeV}},\\
m_{D^{\pm}}=(1.86961\pm 0.00009) {\rm{GeV}}, & f_{D}=(0.2037 \pm 0.0049) {\rm{GeV}} ,\\
m_{a_{0}(980)}=(0.980\pm 0.020) {\rm{GeV}}, & m_{e}=0.5109989461\times 10^{-3} {\rm{GeV}},\\
\tau_{D}=(410.1\pm1.5)\times 10^{-15}s,& \tau_{D^{+}}=(1040\pm7)\times 10^{-15}s \label{inputs}
\end{array}
\end{align}
and
\begin{align}
\begin{array}{ll}
\langle \alpha_s G^2 \rangle = (7.5 \pm 2)\times10^{-2} \ {\rm{GeV}}^4, &\langle g_s^3fG^3 \rangle=(8.2 \pm 1) \times \langle \alpha_s  G^2 \rangle, \\
\langle \bar{u} u\rangle \cong \langle\bar{d} d\rangle \cong -(0.254 \pm 0.015)^3 {\rm{GeV}}^3 ,& g_s^2\langle \bar{u}u\rangle^2=g_s^2 \langle \bar{d}d\rangle^2 =2.693\times 10^{-3} \ {\rm{GeV}}^6\\
\langle g_s\bar{u}\sigma TG u\rangle \cong\langle g_s\bar{d}\sigma T G d\rangle=m_0^2 \langle \bar{u}u\rangle, &
m_0^2=(0.80 \pm 0.02) \ {\rm{GeV}}^2, \label{condensatesparameter}
\end{array}
\end{align}
The condensate parameters are given at the scale $\mu=2$ ${\rm{GeV}}$, which can be run to any required scales with the evolution equations.
\subsection{Decay constants for the scalar mesons $a_0(980)$}
\begin{figure}[t]
\centering
\includegraphics[width=0.42\textwidth]{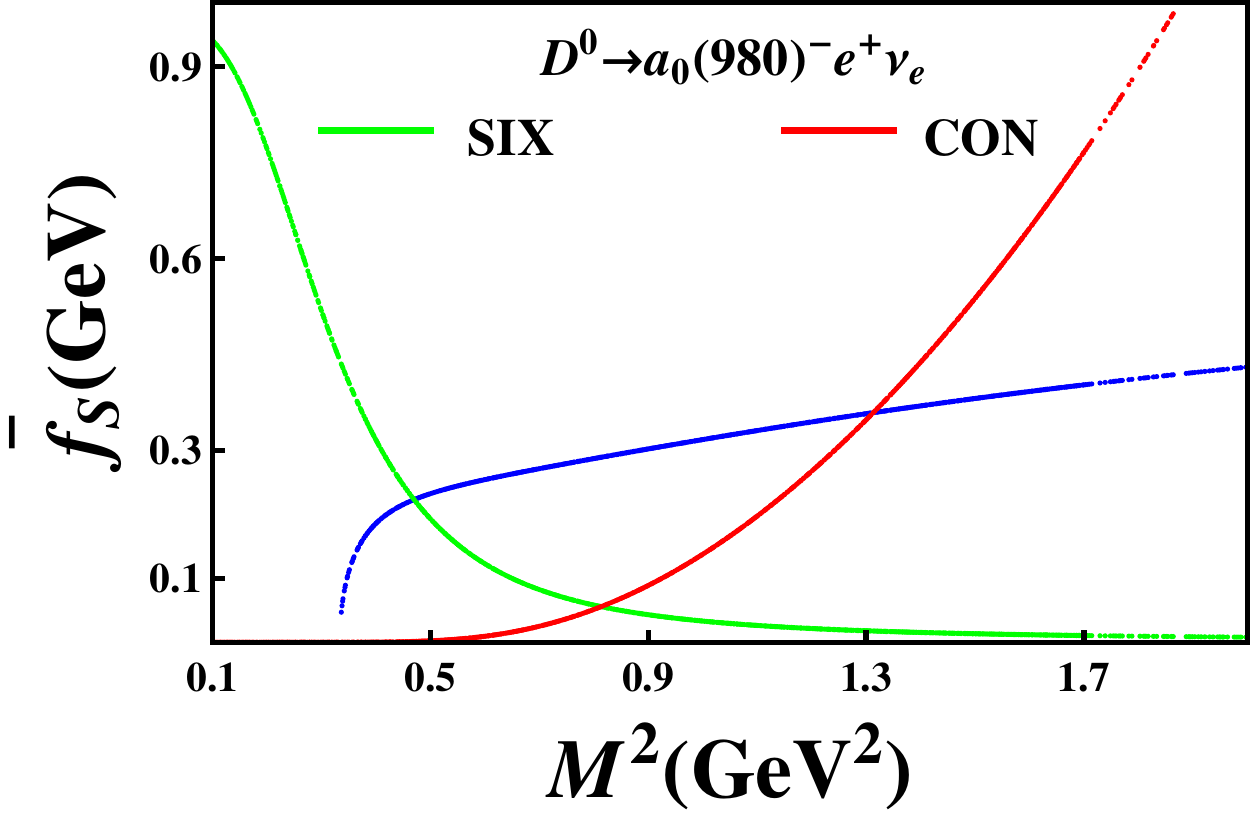}\qquad
\includegraphics[width=0.42\textwidth]{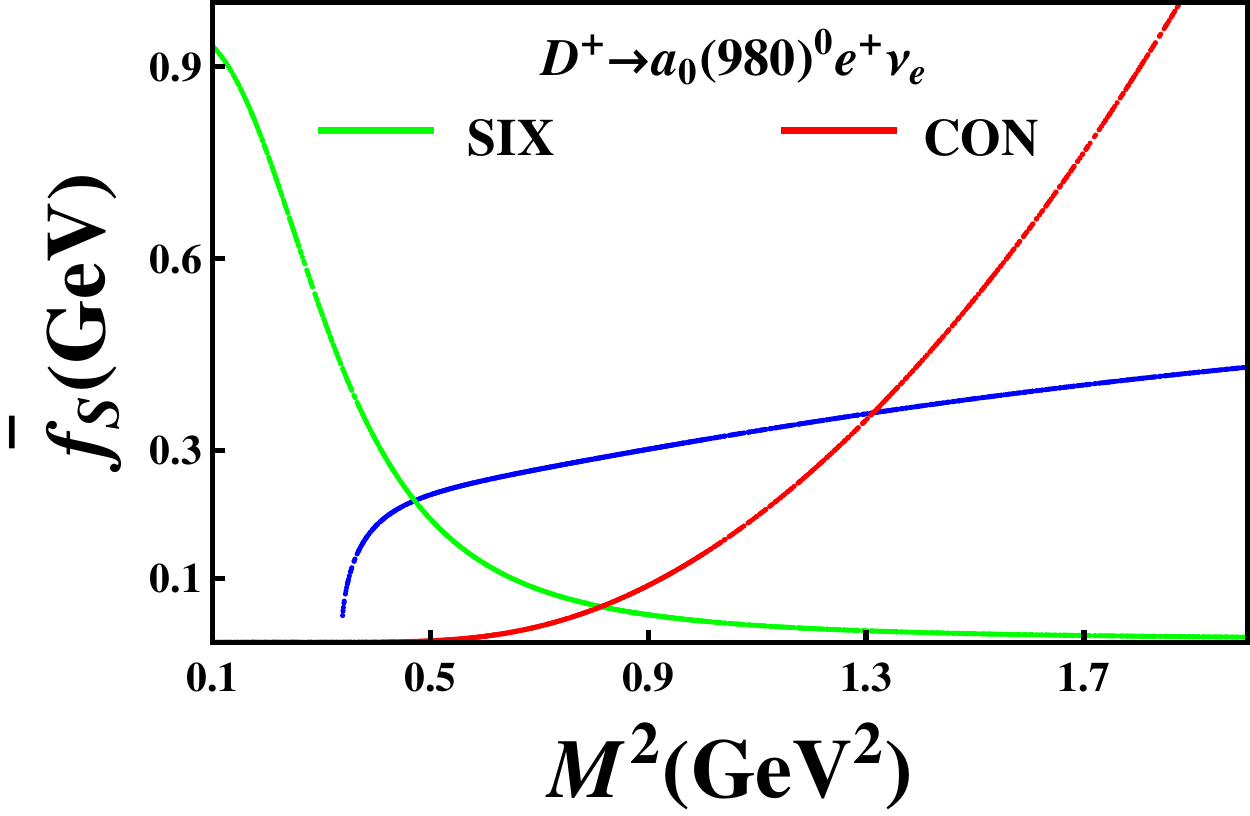}
\caption{\small Decay constants of $a_0(980)$ versus $M^2$. The green line denote the fraction of the dimension-six condensate contribution (SIX), and the red line the fraction of the continuum contribution (CON).}
\label{fsmsqfig}
\end{figure}

Taking the mass of $a_0(980)$ as input and setting $n=0$ in the sum rules (Eqs.\ref{eq:evenmomentssigma},\ref{eq:evenmomentsd} ) for the moment of $\phi_{S}^{s}$ and $\phi_{S}^{\sigma}$, such as $\langle \xi^{0}_s\rangle$, we can calculate the decay constant of $a_0(980)$. The threshold parameters $S_s$ and $S_{\sigma}$ are taken to be $(5.0\pm0.3)\mbox{GeV}^2$ for $a_0(980)$ in the scenario that the scalar mesons are made of two quarks. The Borel window of $a_0(980)$ is determined by the following criteria according to the SVZ sum rule, where the contributions from the dimension-six condensate (SIX) are less than $10\%$ in the total sum rules and the contribution of continuum state (CON) does not exceed $20\%$ of the total dispersive integration. Following these criteria, we can obtain the decay constants of $a_0(980)$ and the corresponding Borel windows at the energy scale 1 GeV
\begin{align}
\bar{f}_{S}=(0.26\sim0.33) {\rm GeV},&& M^2=(0.65\sim1.10){\rm GeV^2},\label{fsbarborel}
\end{align}
The variation of the decay constant $\bar{f}_{S}$ versus the Borel parameter $M^2$ are presented in Fig.~\ref{fsmsqfig}, where one can see
that the decay constant of $a_0(980)$ is in a good stability against the variation of $M^2$ within the Borel window.
\subsection{Moments for the scalar mesons $a_0(980)$}
\begin{figure}[t]
\centering
\includegraphics[width=0.42\textwidth]{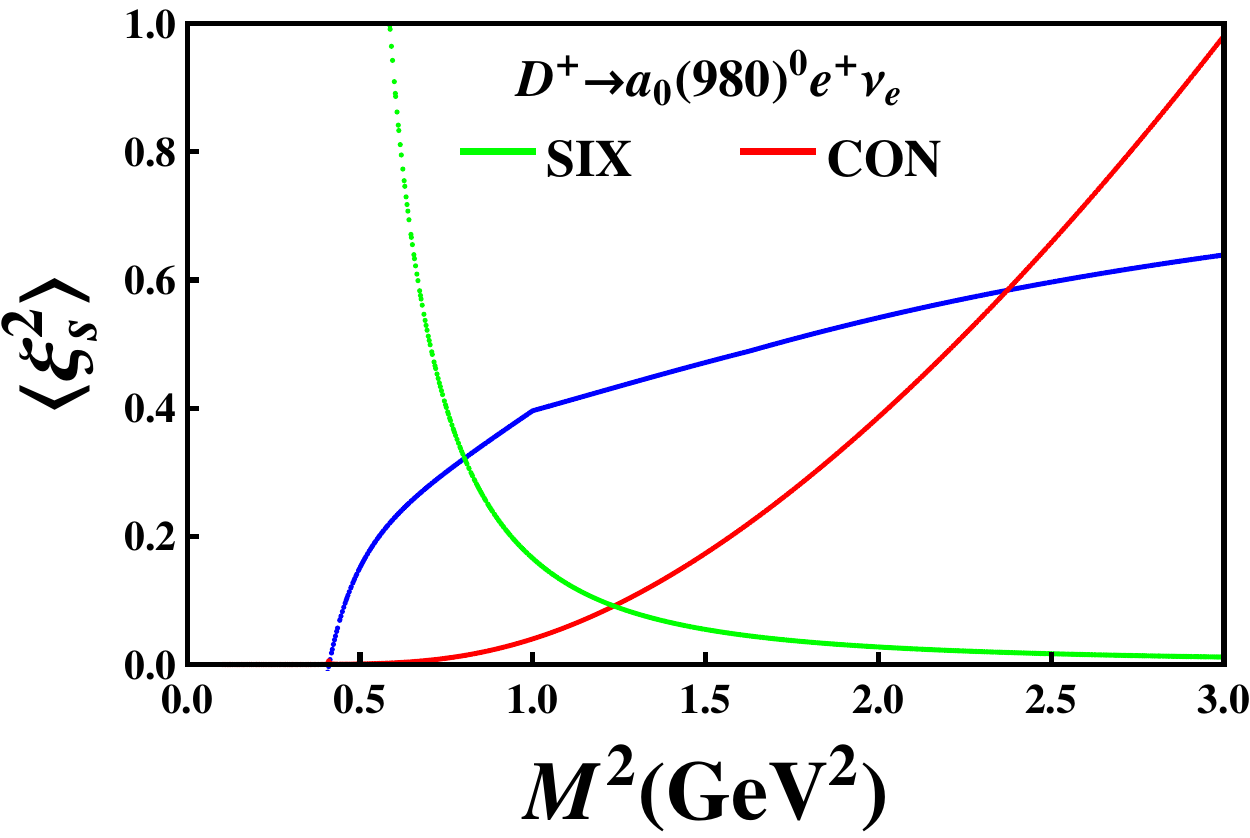}\qquad
\includegraphics[width=0.42\textwidth]{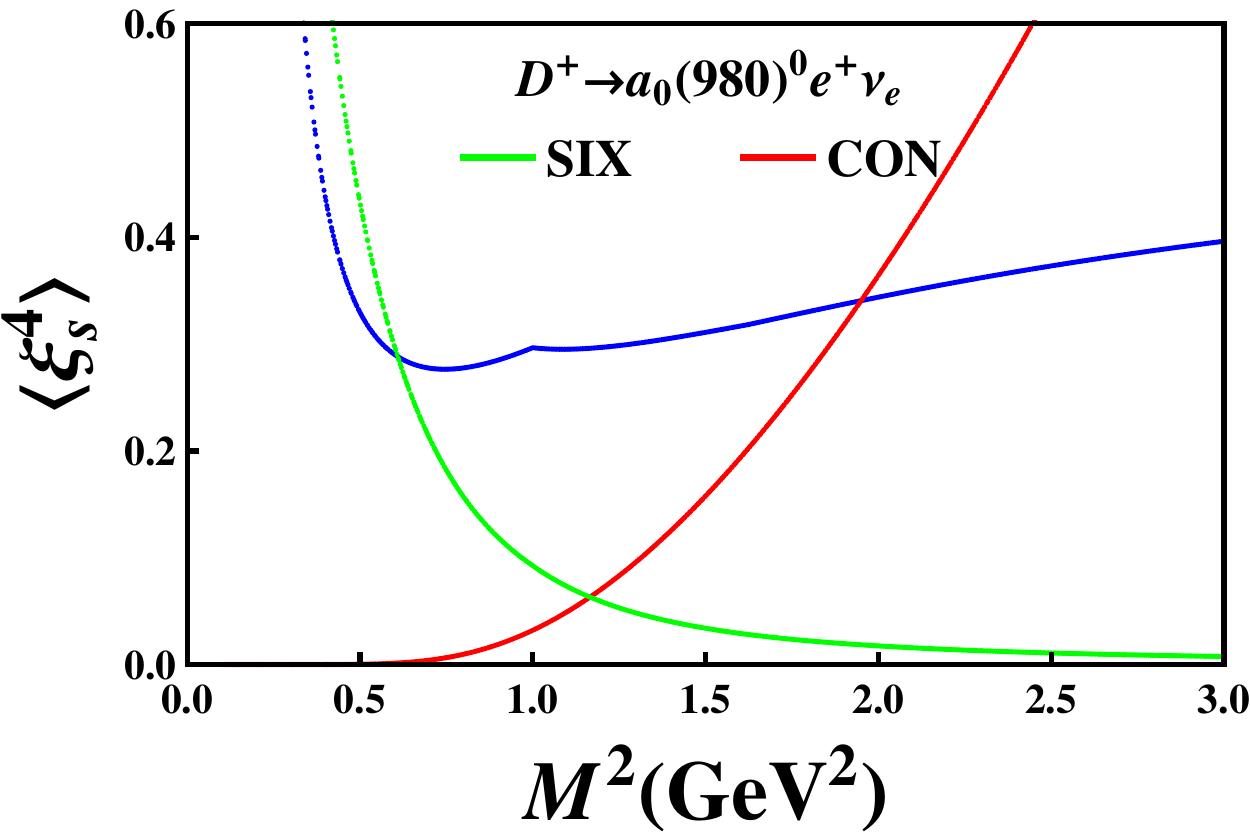}
\includegraphics[width=0.42\textwidth]{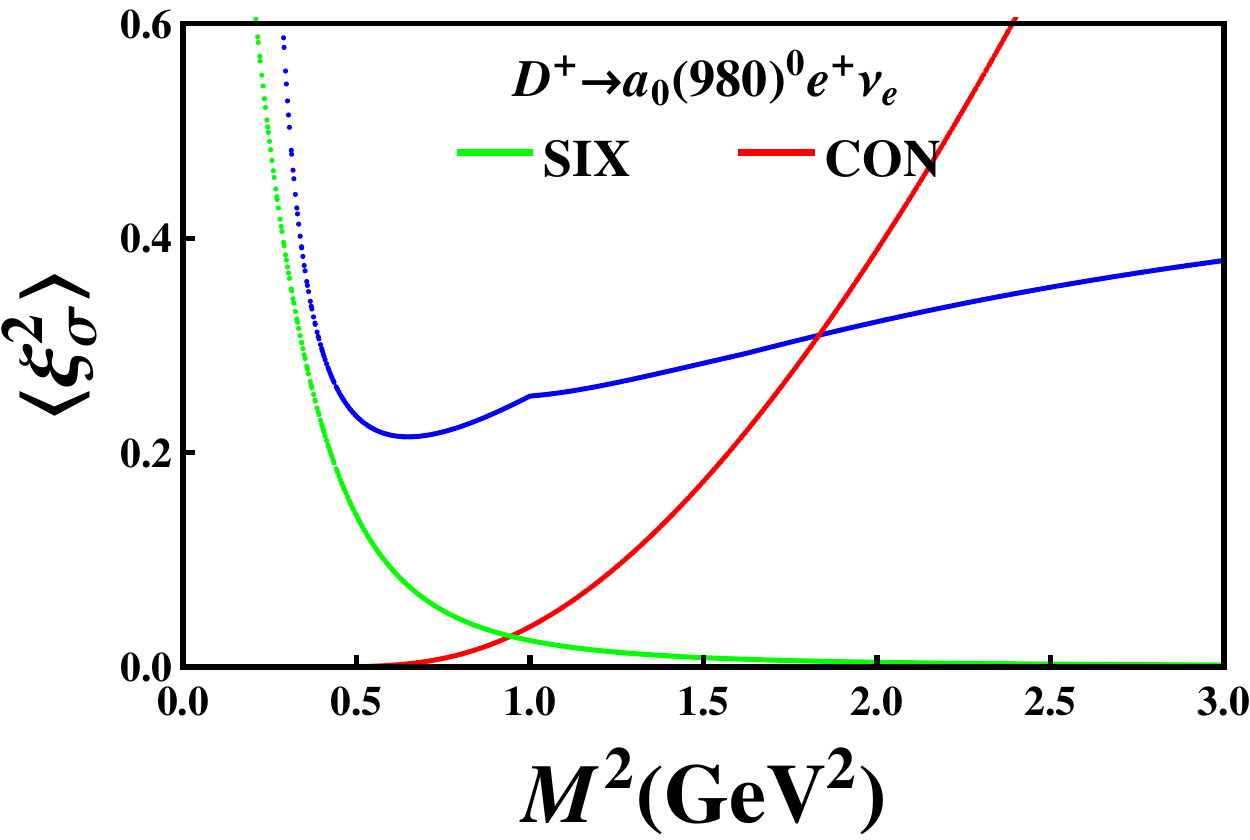}\qquad
\includegraphics[width=0.42\textwidth]{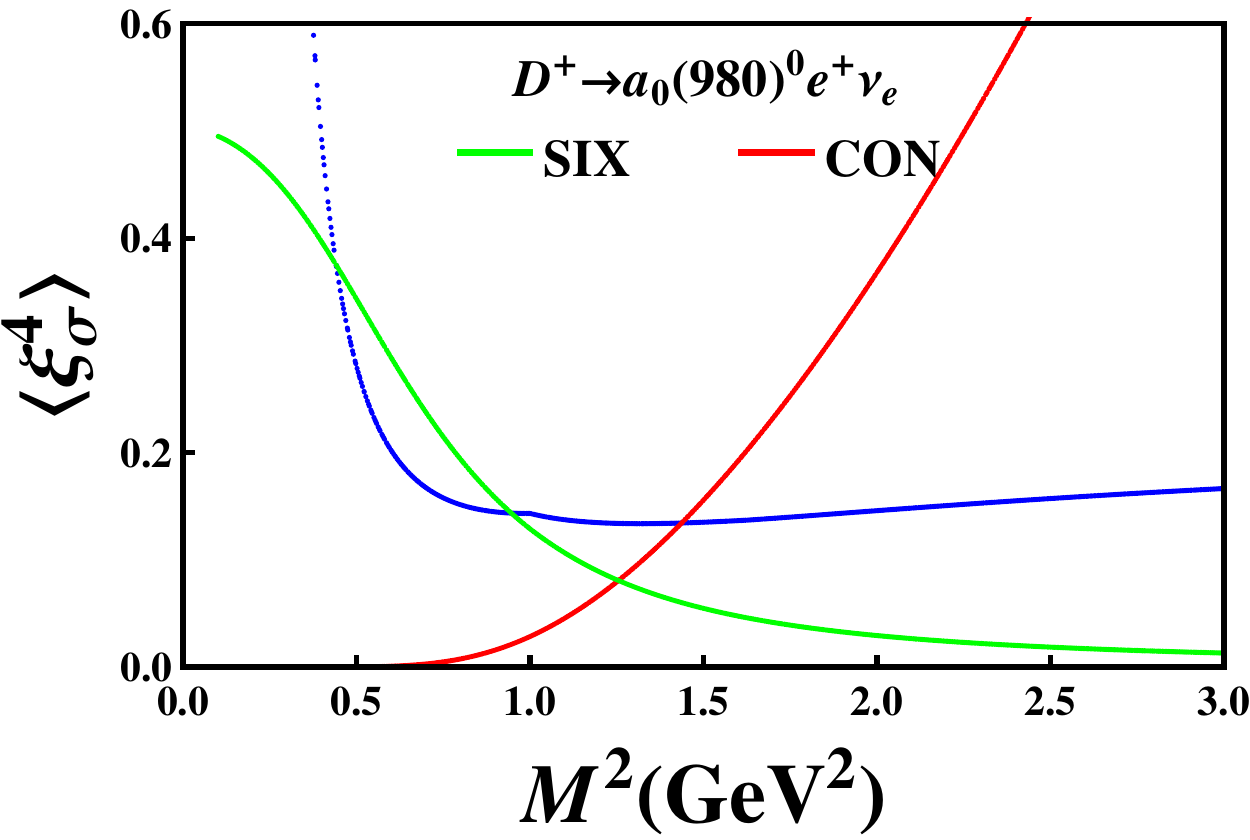}
\caption{\small The moments $\langle \xi_s^{2(4)}\rangle$ and $\langle \xi_{\sigma}^{2(4)}\rangle$ for $a_0^0(980)$ versus $M^2$. The green line denotes the fraction of the contribution of dimension-six condensate (SIX), and the red line the fraction of the continuum contribution (CON).}
\label{momentsmsqazfig}
\end{figure}

\begin{figure}[t]
\centering
\includegraphics[width=0.42\textwidth]{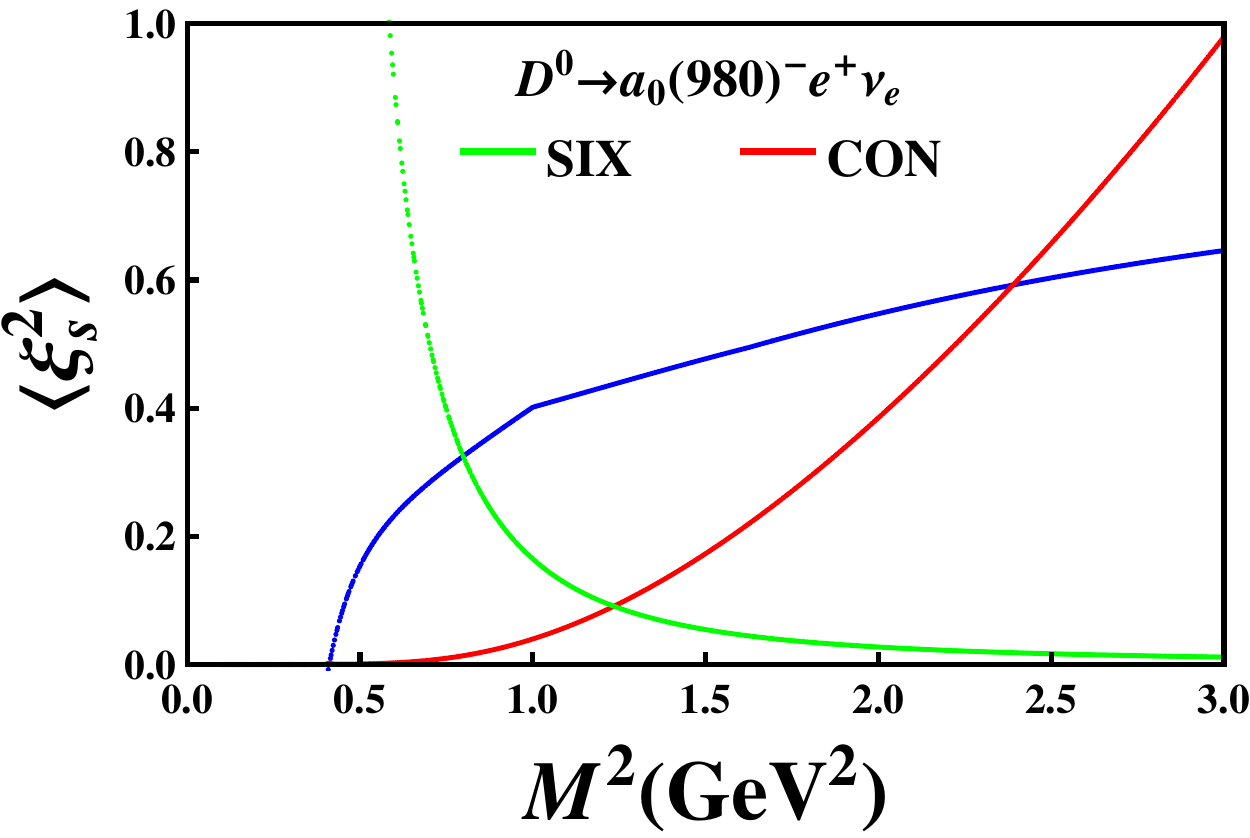}\qquad
\includegraphics[width=0.42\textwidth]{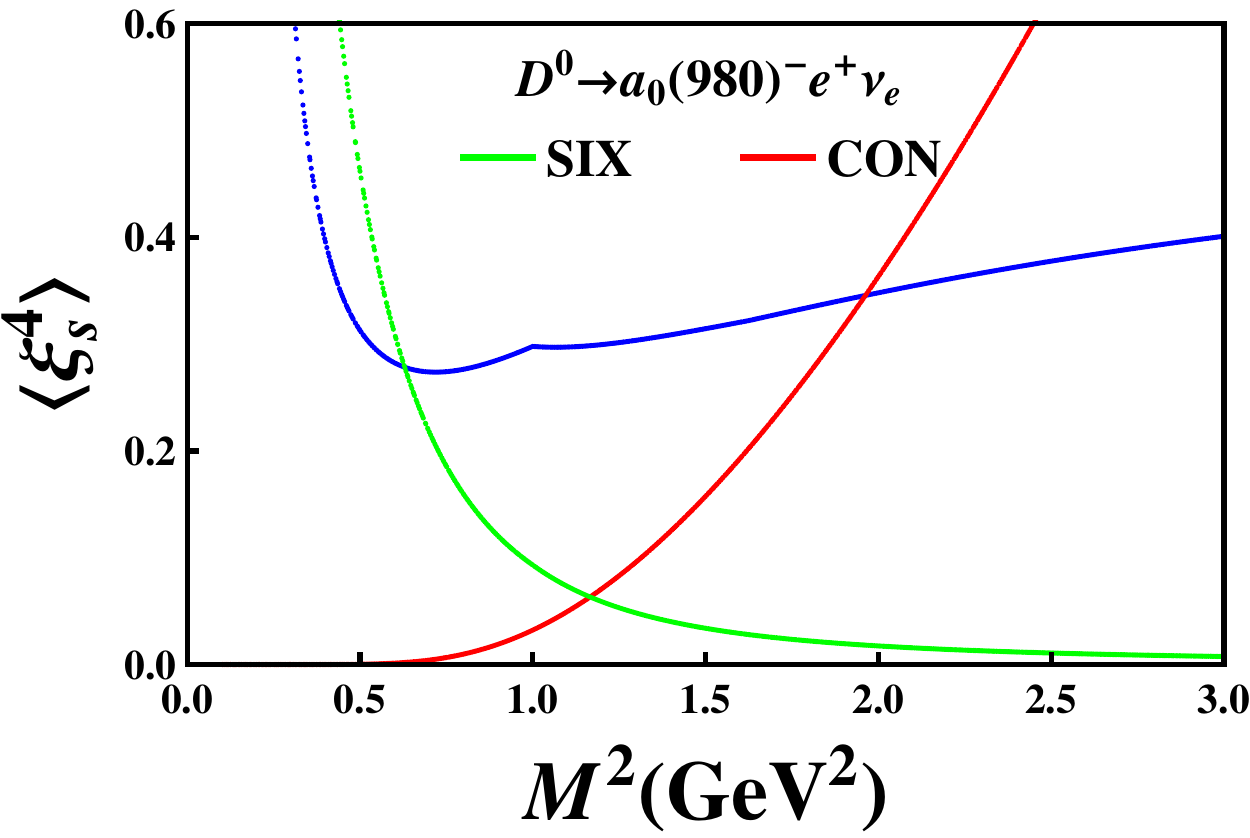}
\includegraphics[width=0.42\textwidth]{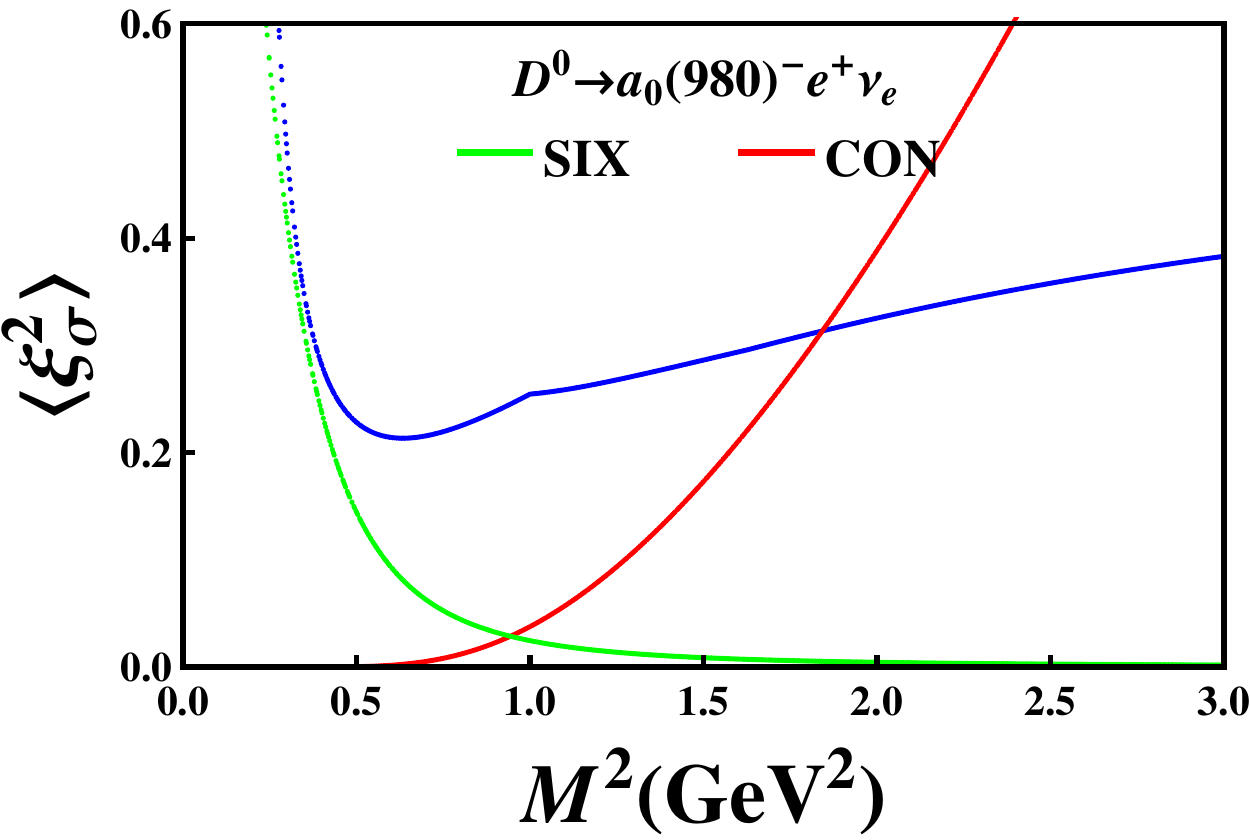}\qquad
\includegraphics[width=0.42\textwidth]{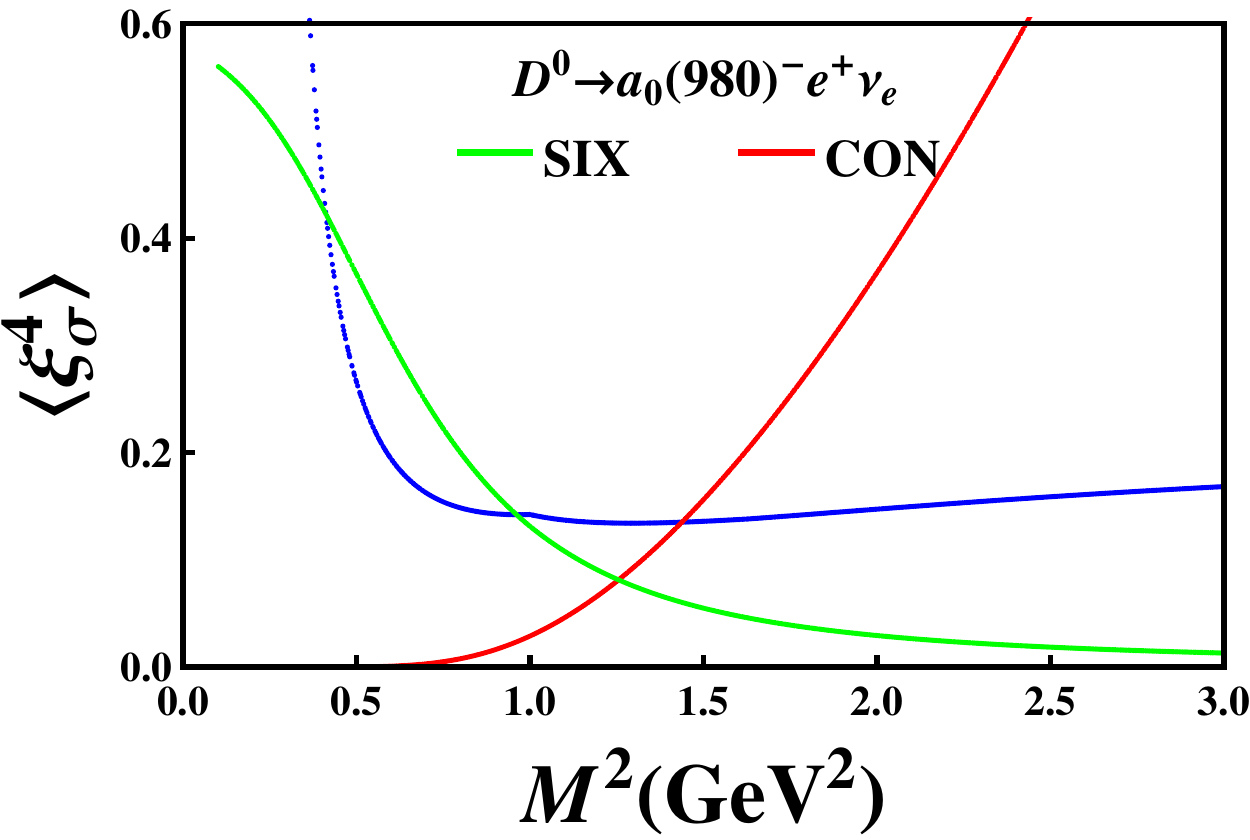}
\caption{\small The moments $\langle \xi_s^{2(4)}\rangle$ and $\langle \xi_{\sigma}^{2(4)}\rangle$ for $a_0^-(980)$ versus $M^2$. The green line denotes the fraction of the contribution of dimension-six condensate (SIX), and the red line the fraction of the continuum contribution (CON).}
\label{momentsmsqapfig}
\end{figure}

Taking the mass and decay constant for $a_0(980)$ as input, we can further calculate the moment of the twist-2 distribution amplitude $\Phi_S$ and the twist-3 distribution amplitudes $\Phi_S^s$ and $\Phi_S^\sigma$. In the two-quark picture, the twist-2 distribution amplitude $\Phi_S(u,\mu)$ for $a_0(980)$ are antisymmetric under the interchange $u\leftrightarrow 1-u$ in the flavor $SU(3)$ limit, so $B_0\left( \mu \right)$, $B_2\left( \mu \right)$ and $B_4\left( \mu \right)$ vanish in that limit. In the following, we will only take into account $B_1\left( \mu \right)$ and $B_3\left( \mu \right)$, which are given in terms of the moments $\langle \xi_{\phi}^1\rangle$ and $\langle \xi_{\phi}^3\rangle$ in Eq. (\ref{relationC}). We use the numerical results for $\langle \xi_{\phi}^1\rangle$ and $\langle \xi_{\phi}^3\rangle$ given in Ref~\cite{Cheng:2005nb}, which is based on the QCDSR method at $\mu=1$ GeV. The values of $\langle \xi_{\phi}^1\rangle$ and $\langle \xi_{\phi}^3\rangle$, together with the corresponding Borel windows, are
\begin{align}
\langle \xi_{\phi}^1\rangle=-0.56\pm 0.05,&&M^2=(1.10\sim1.60){\rm GeV^2},\label{momentphi1}\\
\langle \xi_{\phi}^3\rangle=-0.21\pm 0.03,&&M^2=(1.40\sim1.90){\rm GeV^2},\label{momentphi3}
\end{align}
As for the twist-3 distribution amplitude, we calculated the moments of $a_0(980)$ from the sum rules given in Eqs.(\ref{eq:evenmomentssigma}), (\ref{eq:evenmomentsd}), (\ref{eq:oddmomentsd}), (\ref{eq:oddmomentssigma}). Here, we note that the odd moments for $a_0(980)$ vanish when the conservation of charge parity and isospin symmetry is considered, so we take only into account the first two even moments $\langle \xi_s^{2(4)}\rangle$ and $\langle \xi_{\sigma}^{2(4)}\rangle$ and neglect the odd moments of $a_0(980)$. In order to find the stable Borel window for the sum rules for the moments, we require that the contributions of dimension-six condensate (SIX) are less than $5\%$ in the total sum rules and the continuum contribution (CON) does not exceed $20\%$ of the total dispersive integration. The moments for $a_0(980)$, the ratio of contribution from the dimension-six condensate and the ratio of the continuum contribution versus $M^2$ are shown in Figs.~\ref{momentsmsqazfig} and Fig.~\ref{momentsmsqapfig}.
From these figures, we can see that the moments within the Borel windows are stable and the values of $\langle \xi_s^{2(4)}\rangle$ and $\langle \xi_{\sigma}^{2(4)}\rangle$ within their stable Borel window at the energy scale 1 GeV are
\begin{align}
\langle \xi_{\sigma,a_0^0}^2\rangle=0.25\pm0.03,&&M^2=(0.77\sim1.57){\rm GeV^2},\label{azmomentsigma1}\\
\langle \xi_{\sigma,a_0^0}^4\rangle=0.14,&&M^2=(1.56\sim1.62){\rm GeV^2},\label{azmomentsigma3}\\
\langle \xi_{s,a_0^0}^2\rangle=0.48,&&M^2=(1.56\sim1.57){\rm GeV^2},\label{azmomentsd1}\\
\langle \xi_{s,a_0^0}^4\rangle=0.31\pm0.01,&&M^2=(1.28\sim1.62){\rm GeV^2},\label{azmomentsd3}
\end{align}
and
\begin{align}
\langle \xi_{\sigma,a_0^-}^2\rangle=0.26\pm0.04,&&M^2=(0.77\sim1.57){\rm GeV^2},\label{apmomentsigma1}\\
\langle \xi_{\sigma,a_0^-}^4\rangle=0.14,&&M^2=(1.57\sim1.62){\rm GeV^2},\label{apmomentsigma3}\\
\langle \xi_{s,a_0^-}^2\rangle=0.49\pm0.01 ,&&M^2=(1.56\sim1.57){\rm GeV^2},\label{apmomentsd1}\\
\langle \xi_{s,a_0^-}^4\rangle=0.31\pm0.01,&&M^2=(1.29\sim1.62){\rm GeV^2},\label{apmomentsd3}
\end{align}
where $\xi_{\sigma,a_0^0}^{2(4)}$ and $\xi_{s,a_0^0}^{2(4)} $denote the moments for $a_0^0(980)$, while $\xi_{\sigma,a_0^-}^{2(4)}$ and $\xi_{s,a_0^-}^{2(4)}$ the moments for $a_0^-(980)$.
\subsection{The form factors of the $D\rightarrow a_0(980)$ transition}
\begin{figure}[t]
\centering
\includegraphics[width=0.42\textwidth]{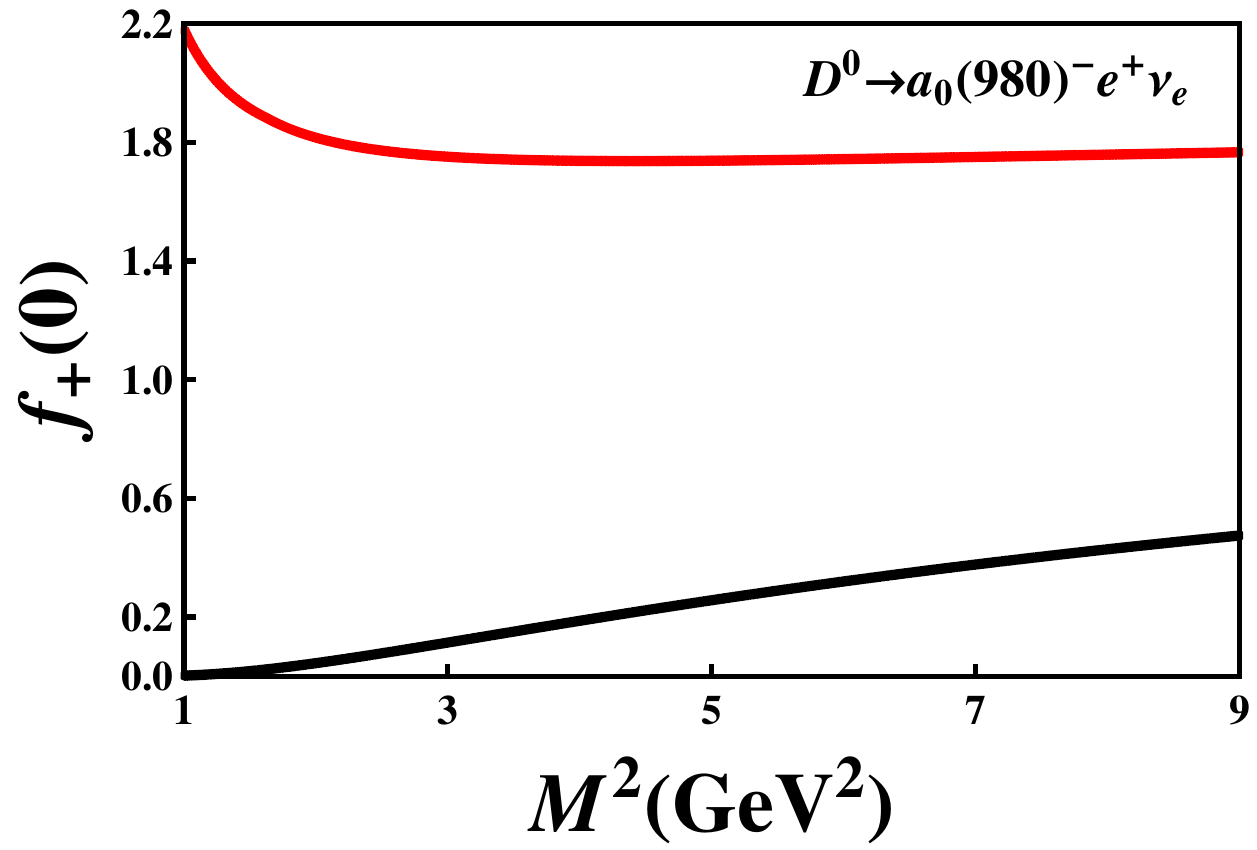}\qquad
\includegraphics[width=0.42\textwidth]{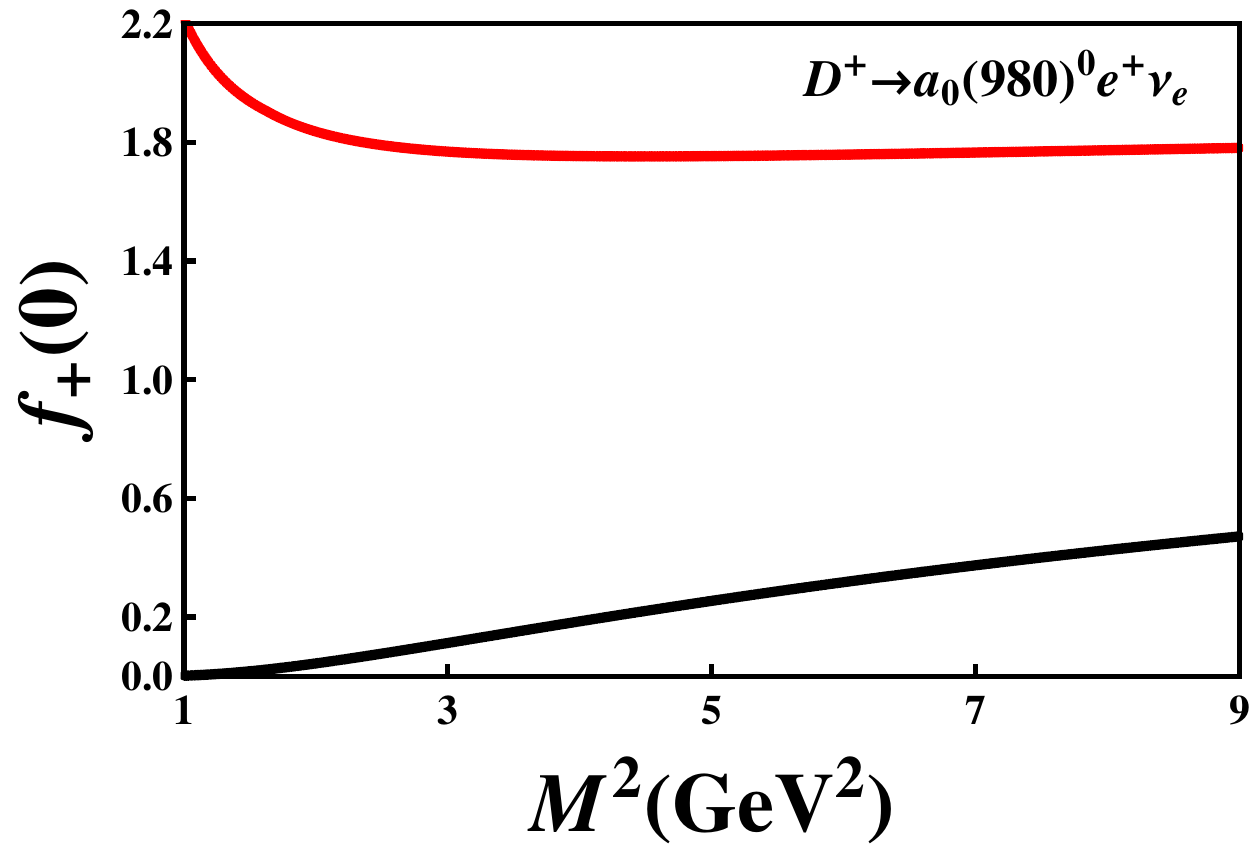}
\caption{\small The LCSR for form factor $f_+ (0)$ versus $M^2$. The red line denotes the value of $f_+ (0)$, and the black line the ratio of the higher excited resonances and continuum states contribution.}
\label{figfzmds}
\end{figure}

With the sum rules for the form factors $f_{\pm}(q^2)$ in Eq.~(\ref{fplus}) and Eq.~(\ref{fminus}), we can proceed the numerical calculation for the form factors. The threshold parameter $s_0$, which is corresponds to the mass of the lowest pseudoscalar D meson, can be estimated in several effective scenarios~\cite{Chernyak:1990ag,Navarra:2000ji,Dosch:2002rh,Matheus:2002nq,Bracco:2004rx,Lucha:2009uy}. We adopt $s_0=(5.37\pm0.13){\rm GeV}^2$ corresponding to D channel. In order to determine the range of Borel parameter $M^2$, we consider the LCSR for form factors $f_{+}(0)$, and require that the contributions of the higher excited resonances and continuum states do not exceed $20\%$ and the value of $f_{+}(0)$ mildly varies with respect to the Borel parameter. In view of these considerations, the range of the Borel parameter $M^2$ is determined as $2.51{\rm GeV^2}\leq M^2\leq 4.18{\rm GeV^2}$ for the decay channel $D^0\rightarrow a_0^- (980) e^+ \nu_e$, and $2.55{\rm GeV^2}\leq M^2\leq 4.21{\rm GeV^2}$ for the decay channel $D^+\rightarrow a_0^0 (980) e^+ \nu_e$. The Borel parameter dependence of the LCSR for form factor $f_{+}(0)$ is shown in Fig.~\ref{figfzmds}.

Next, we investigate the $q^2$-dependence of the form factors $f_{+}(q^2)$ and $f_{-}(q^2)$ based on the sum rules in Eqs.(\ref{fplus}) and (\ref{fminus}). We adopt the double-pole form to parameterize the form factors $f_{i} (q^2) (i=+,-)$
\begin{align}
f_{i}(q^2)={f_i(0) \over 1-a_i q^2/m_{D}^{2}+b_i
q^4/m_{D}^{4}}, \label{double-pole model of form factors}
\end{align}
in the kinematical region $m_e^2\le q^2\le (m_D-m_{a_{0}(980)})^2$. Here, $a_i$ and $b_i$ are the non-perturbative parameters, which can be fixed by the values of form factors at the small and intermediate $q^2$ in the LCSR approach.
The numerical results for the parameters $f_{i}$, $a_i$ and $b_i$ are shown in Table~\ref{dzfpfzparaval} and Table~\ref{dpfpfzparaval},
where the theoretical uncertainties are caused by varying the Borel parameter $M$, the threshold value parameter $s_0$, the $c$ quark
mass, the decay constants and masses of the involved mesons and the Gengenbauer moments for the twist-2 and twist-3 LCDAs of $a_0 (980)$.
\begin{table}[t]
\begin{center}
\caption{\label{dzfpfzparaval} \small Numerical results for the parameters $f_{i}$, $a_i$ and $b_i$ involved in the double-pole fit of form factors for $D^0\rightarrow a_0^- (980) e^+ \nu_e$. }
\vspace{0.2cm}
\doublerulesep 0.8pt \tabcolsep 0.18in
\begin{tabular}{c|c|c|c}
\hline\hline
         &  $f_{i}(0)$                 &  $a_i$                      &   $b_i$                     \\
\hline
$f_+$    &  $1.75_{-0.27}^{+0.26}$  &  $0.54$  &   $0.91_{-0.09}^{+0.17}$  \\
\hline
$f_-$    &  $0.31\pm 0.13$  &  $1.14_{-1.11}^{+2.46}$  &   ${1.70}_{-3.97}^{+3.00}$  \\
\hline\hline
\end{tabular}
\end{center}
\end{table}
\begin{table}[t]
\begin{center}
\caption{\label{dpfpfzparaval} \small Numerical results for the parameters $f_{i}$, $a_i$ and $b_i$ involved in the double-pole fit of form factors for $D^+\rightarrow a_0^0 (980) e^+ \nu_e$. }
\vspace{0.2cm}
\doublerulesep 0.8pt \tabcolsep 0.18in
\begin{tabular}{c|c|c|c}
\hline\hline
         &  $f_{i}(0)$                 &  $a_i$                      &   $b_i$                     \\
\hline
$f_+$    &  $1.76\pm{0.26}$  &  $0.55\pm{0.01}$  &   $0.94_{-0.08}^{+0.16}$  \\
\hline
$f_-$    &  $0.31\pm{0.13}$  &  $1.23_{-1.12}^{+2.36}$  &   ${-1.55}_{-0.92}^{+6.30}$  \\
\hline\hline
\end{tabular}
\end{center}
\end{table}
\subsection{The decay rates for the $D\rightarrow a_0(980)$ semileptonic decays}
\begin{figure}[t]
\centering
\includegraphics[width=0.42\textwidth]{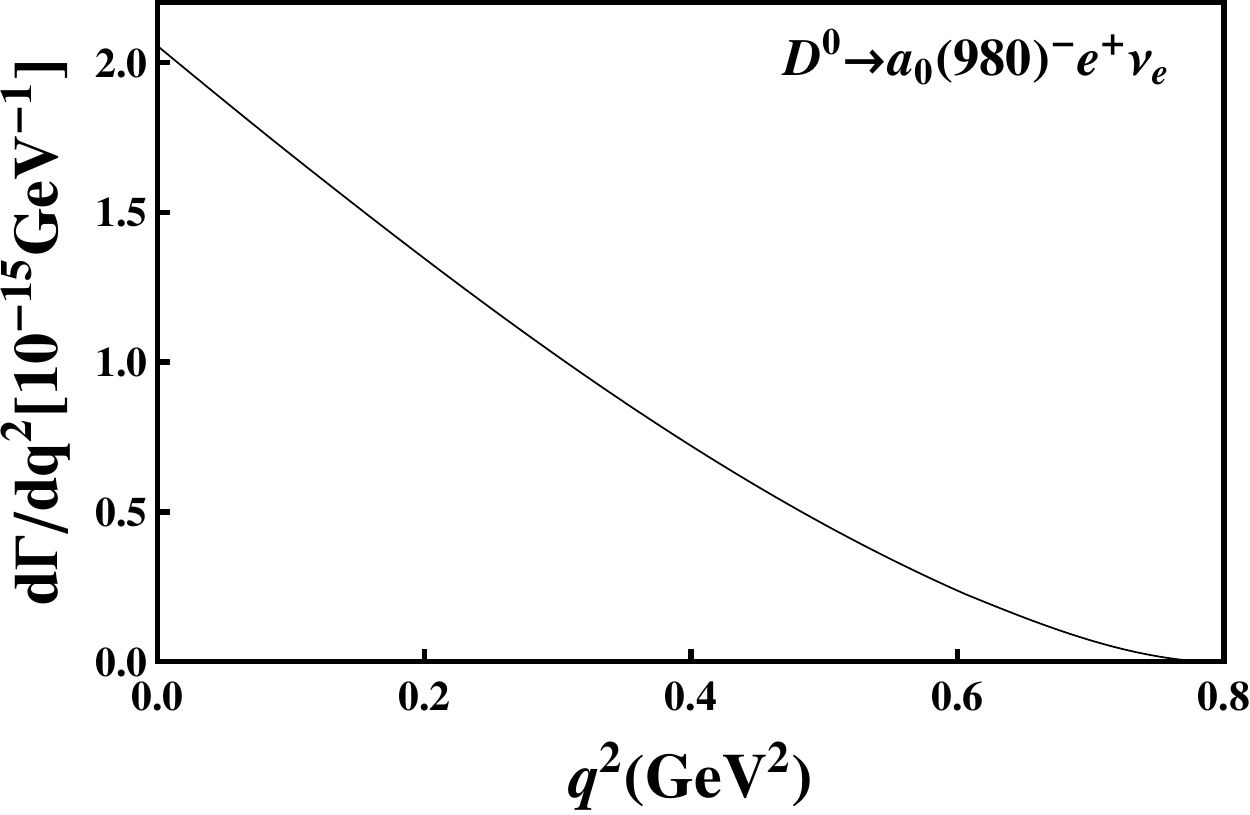}\qquad
\includegraphics[width=0.42\textwidth]{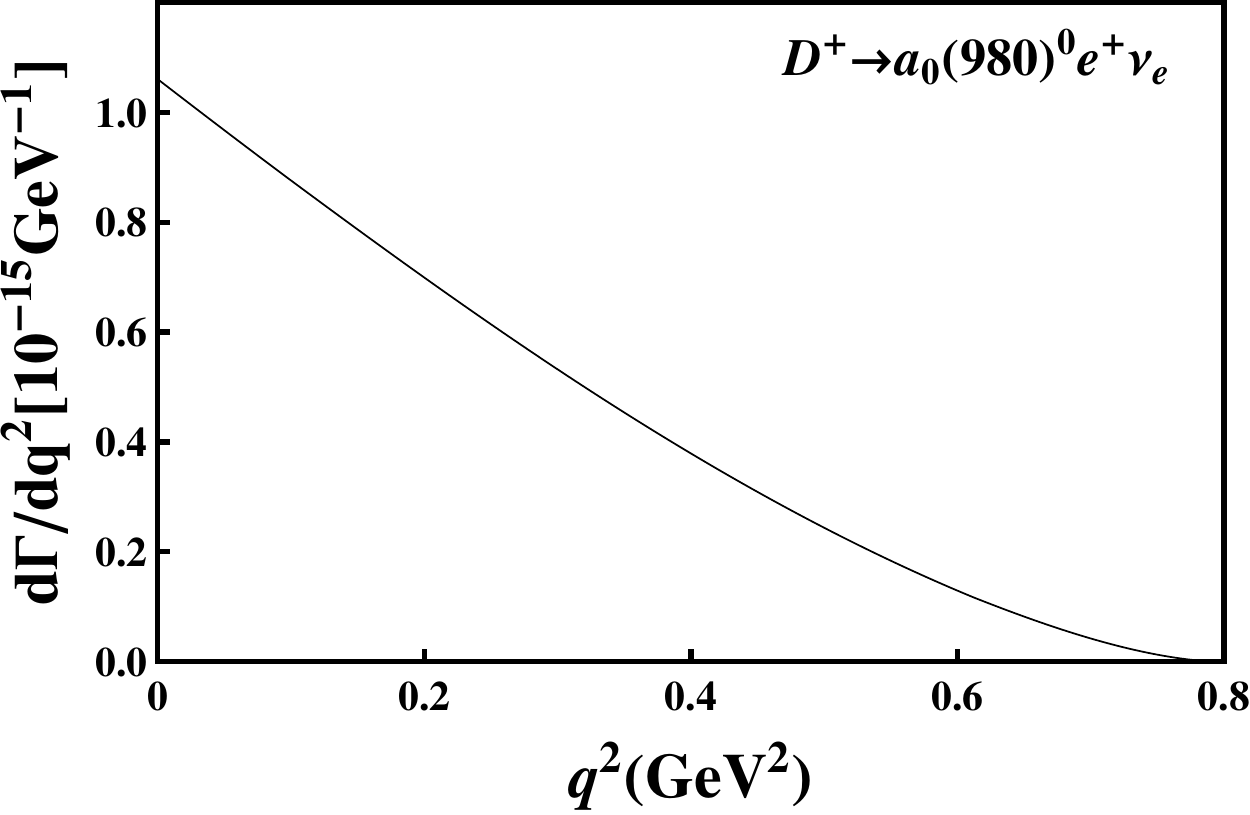}
\caption{\small The differential decay width for $D^0\rightarrow a_0^- (980) e^+ \nu_e$ and $D^+\rightarrow a_0^0 (980) e^+ \nu_e$.}
\label{figdifferentrate}
\end{figure}

Using Eq.(\ref{eq:widthsemi}) and the transition form factors obtained in the LCSR, we are in a position to calculate the differential decay rates and the branching ratios for the $D\rightarrow a_0(980)$ semileptonic decays.
In Eq.(\ref{eq:widthsemi}), the terms involving $f_{+} (q^2) f_{-} (q^2)$ and $(f_{-} (q^2))^2$ are suppressed by the small mass of electron, so the differential decay rates and the branching ratios are dominated by the term of $(f_{+} (q^2))^2$. We present the distributions of the differential decay rates for $D^0\rightarrow a_0^- (980) e^+ \nu_e$ and $D^+\rightarrow a_0^0 (980) e^+ \nu_e$ in the kinematic region $m_e^2\le q^2\le (m_D-m_{a_{0}(980)})^2$ in Fig.~\ref{figdifferentrate}.
Furthermore, we can obtain the branching ratios of $D^0\rightarrow a_0^- (980) e^+ \nu_e$ and $D^+\rightarrow a_0^0 (980) e^+ \nu_e$ by performing the integration of the differential decay rates over the momentum transfer squared $q^2$. The results are
\begin{align}
 {\mathcal B}(D^0\rightarrow a_0^- (980) e^+ \nu_e)=(4.08^{+1.37}_{-1.22})\times 10^{-4},\\
{\mathcal B}(D^+\rightarrow a_0^0 (980) e^+ \nu_e)=(5.40^{+1.78}_{-1.59})\times 10^{-4},\label{branchingratio}
\end{align}
As for $ a_0 (980)$, the dominant decay modes are $\eta \pi$ and $K\bar{K}$. By applying the averaged value in the Particle Data Group (PDG) $\Gamma(a_0(980)\rightarrow K\bar{K})/\Gamma(a_0(980)\rightarrow \eta \pi)=0.183\pm0.024$~\cite{Olive:2016xmw}, one can obtain~\cite{Cheng:2005nb,Cheng:2013fba}
\begin{align}
 {\mathcal B}( a_0 (980)\rightarrow \eta\pi)=0.845\pm0.017,\label{branchingratioofa980}
\end{align}
so the branching ratios  ${\mathcal B}(D\rightarrow a_0 (980) e^+ \nu_e;a_0(980)\rightarrow\eta\pi)$ will have values above $10^{-4}$, which can be observed using the current data samples, such as the sample of $D\bar{D}$ pairs at CLEO-c and the sample of $D\bar{D}$ pairs at BESIII. If these decay channels have no signal to be observed or the measurements of the branching ratios disagree with the above predictions, the 2-quark picture for $a_0 (980)$ will be disfavored.
\section{Conclusion}
In this work, we study the semileptonic decays of $D^0\rightarrow a_0^- (980) e^+ \nu_e$ and $D^+\rightarrow a_0^0 (980) e^+ \nu_e$ in the LCSR approach, where the scalar $a_0(980)$ is assumed as $q\bar{q}$ state. We calculate the form factors responsible for these decays up to twist-3 distribution amplitudes for the scalar meson. The differential decay rates and branching ratios of $D^0\rightarrow a_0^- (980) e^+ \nu_e$ and $D^+\rightarrow a_0^0 (980) e^+ \nu_e$ are obtained. We find that these decay channels can be hopefully observed in experiment, which might be beneficial to identify the inner structures of $a_0(980)$.
\section*{Acknowledgements}
This work is supported in part by the National Natural
Science Foundation of China under Contracts Nos.~11647067, 11335009, 11125525,11675137, the Joint Large-Scale Scientific Facility Funds of the NSFC and CAS
under Contract No.~U1532257, CAS under Contract No.~QYZDJ-SSW-SLH003, and the National Key Basic Research Program of China under Contract
No.~2015CB856700.  X.~Cheng and Y.~Xu are supported by Nanhu Scholars Program of XYNU.

\begin{appendix}

\section{The sum rules for the moments of $\Phi _{S}^{s}$ and $\Phi _{S}^{\sigma}$}
\label{appendix:1}
The sum rule for the even moments of $\phi_{S}^{\sigma}$ up to dimension-six condensates can be expressed as~\cite{Han:2013zg}:
\begin{align}
&-\frac{1}{3}{m_{a_0 (980)}^2} \bar{f}_S^2 e^{-{m_{a_0 (980)}^2}/M^2}\langle \xi^{2n}_{\sigma}\rangle \nonumber \\
=& \frac{3}{4\pi^2}\int^1_0dx(2x-1)^{2n}M^4x(x-1)e^{-\frac{m_{12}^2}{M^{2}x(1-x)}} \nonumber \\
&-\frac{3}{4 \pi^2}\int^1_0dx (2x-1)^{2n}\left\{x(x-1)\left(1+\frac{S_{\sigma}}
{ M^2}\right) +\frac{m_{12}^2}{M^2}\right\} M^4 e^{-{S_{\sigma}/M^2}} \nonumber \\
&-\langle \alpha_s G^2\rangle \int^1_0dx\frac{(2x-1)^{2n}}{24\pi}
\left\{ 1-\frac{2m_1m_2}{M^2x(1-x)} \right\}e^{-\frac{m_{12}^2}
{M^{2}x(1-x)}}\nonumber \\
&+\langle g_s^3fG^3\rangle \frac{m_1m_2}{24\pi^2}\int^1_0dx
\frac{(2x-1)^{2n}}{2M^4x(1-x)}e^{-\frac{m_{12}^2}{M^2x(1-x)}}\nonumber \\
&+\bigg \{\langle \bar{q}_1q_1\rangle \frac{-1}{6} \left[ 3m_1+\frac{(4n+1)m_1^3}
{M^2}+\frac{m_1^3m_2^2}{M^4}\right] e^{-m_2^2/M^2}\nonumber \\
&+\langle g_s \bar{q}_1 \sigma TGq_1\rangle \bigg [
\frac{(16n+1)m_1+6m_2}{36M^2}+\frac{m_1m_2^2}{9M^4}\bigg ]e^{-m_2^2/M^2}\nonumber \\
&+\frac{g_s^2\langle \bar{q}_1q_1\rangle^2}{81}\left[ \frac{-4n+5}{M^2}+\frac{2m_2^2}{M^4}\right]
 e^{-m_2^2/M^2} +\left[q_1 \leftrightarrow q_2, m_1 \leftrightarrow m_2\right] \bigg \}
\label{eq:evenmomentssigma},
\end{align}
The sum rule for the scalar density even moments of $\phi_{S}^{s}$ up to dimension-six condensates is
\begin{align}
&-{m_{a_0 (980)}^2} \bar{f}_S^2 e^{-{m_{a_0 (980)}^2}/M^2}\langle \xi^{2n}_s\rangle \nonumber \\
&= + \frac{3}{ 4 \pi^2}\int^{1}_0(2x-1)^{2n}\left [-(2n+3)x(1-x)+\frac{m_1 m_2-m_{12}^2}
{ M^2}\right] M^4e^{-\frac{m_{12}^2 }{ M^{2}x(1-x)}}dx \nonumber  \\
&-\frac{3}{ 4 \pi^2}\int^{1}_0 (2x-1)^{2n}\left[-(2n+3)x(1-x)\left(1+\frac{S_s}{ M^2}\right)
+\frac{2(n+1)m_{12}^2 + m_1 m_2}{ M^2}\right] M^4e^{-{S_s/M^2}}dx \nonumber \\
&+\langle \alpha_s G^2\rangle\int^{1}_0(2x-1)^{2n}\frac{1}{ 8\pi}\left[
-(2n+1)+\frac{2m_1m_2-m_{12}^2} {M^2 x(1-x)}\right] e^{-\frac{m_{12}^2 }{ M^{2}x(1-x)}}dx \nonumber  \\
&+\langle g_s^3fG^3\rangle \frac{n(2n-1)}{48\pi^2}\int^{1}_0(2x-1)^{2n-2} \frac{1}{M^2}e^{-\frac{m_{12}^2 }{ M^{2}x(1-x)}}dx  \nonumber  \\
&+\bigg \{ -\langle \bar{q}_1 q_1\rangle \left[ \frac{2m_2+(2n+1)m_1}{2}\right. \nonumber  \\
&\hspace{1cm}+\frac{2n(4n+1)m_1^3+12nm_1^2m_2+3m_1m_2^2}{6M^2} \left.+\frac{m_1^2m_2^2(2nm_1+m_2)} {2M^4}+\frac{m_1^3m_2^4}{6M^6}\right]e^{-m_2^2/M^2}  \nonumber  \\
&+\langle g_s\bar{q}_1\sigma TGq_1\rangle \bigg [ \frac{9(2n-1)m_2
+n(16n-5)m_1}{18M^2}+\frac{((8n-3)m_1+3m_2)m_2^2}{12M^4}+\frac{m_1m_2^4}
{9M^6}\bigg ]e^{-m_2^2/M^2}\nonumber  \\
&+g_s^2 \langle \bar{q}_1q_1\rangle^2 \left[\frac{-8n^2-14n+12}
{81M^2}-\frac{m_2^2}{27M^4}\right]e^{-m_2^2/M^2}\nonumber  \\
& + g_s^2 \langle \bar{q}_1q_1\rangle \langle \bar{q}_2q_2 \rangle \frac{4}{9}\left[\frac{2e^{-m_2^2/M^2}}{M^2}+\frac{1}{m_2^2}
(e^{-m_2^2/M^2}-1)\right] \nonumber \\
&+g_s^2 \langle \bar{q}_1q_1\rangle \langle \bar{q}_2q_2\rangle
 \frac{2}{9}\frac{1}{m_2^2-m_1^2}\left[(e^{-m_1^2/M^2}-e^{-m_2^2/M^2}) +2m_1 m_2 \left(\frac{e^{-m_1^2/M^2}-1}{m_1^2}-\frac{e^{-m_2^2/M^2}-1}{m_2^2}\right)\right]\nonumber \\
& +\left[q_1 \leftrightarrow q_2, m_1 \leftrightarrow m_2 \right] \bigg \} \label{eq:evenmomentsd},
\end{align}
Here, the zeroth moments are normalized to one, so we can ontain $\langle \xi^{0}_s\rangle=\langle \xi^{0}_{\sigma}\rangle=1$.
The sum rule for the odd moments of $\phi_{S}^{s}$ up to dimension-six condensates is given by
\begin{align}
&-{m_{a_0 (980)}^2} \bar{f}_S^2 e^{-{m_{a_0 (980)}^2}/M^2}\langle \xi^{2n+1}_s\rangle \nonumber \\
&= \frac{3}{ 4 \pi^2}\int^{1}_0(2x-1)^{2n+1}\left[-2(n+2)x(1-x)+\frac{m_1m_2-m_{12}^2}
{ M^2}\right] M^4e^{-\frac{m_{12}^2 }{ M^{2}x(1-x)}}dx \nonumber \\
&-\frac{3}{ 4 \pi^2}\int^{1}_0 (2x-1)^{2n+1}\left[-2(n+2)x(1-x)\left(1+\frac{S_s}{ M^2}\right) +\frac{(2n+3)m_{12}^2 + m_1 m_2} { M^2}\right] M^4 e^{-{S_s/M^2}}dx \nonumber \\
&+\langle \alpha_s G^2\rangle\int^{1}_0(2x-1)^{2n+1}\frac{1}{ 8\pi}\left[
-2(n+1)+\frac{2m_1m_2-m_{12}^2}{M^2 x(1-x)}\right] e^{-\frac{m_{12}^2}{ M^{2}x(1-x)}}dx \nonumber  \\
&+\langle g_s^3fG^3 \rangle \frac{2n(2n+1)}{96\pi^2}\int^{1}_0(2x-1)^{2n-1}
\frac{1}{M^2}e^{-\frac{m_{12}^2}{ M^{2}x(1-x)}}dx  \nonumber  \\
&+\bigg \{-\langle \bar{q}_1q_1\rangle \bigg [ \frac{2m_2+2(n+1)m_1}{2}+ \frac{(2n+1)(4n+3)m_1^3 +6(2n+1)m_1^2m_2+3m_1m_2^2}{6M^2}\nonumber \\
&\hspace{3.8cm}+\frac{m_1^2m_2^2((2n+1)m_1+m_2)} {2M^4}+\frac{m_1^3m_2^4}{6M^6}\bigg ]e^{-m_2^2/M^2}\nonumber \\
& +\langle g_s\bar{q}_1\sigma T Gq_1\rangle \left[ \frac{36nm_2 +(2n+1)(16n+3)m_1}{36M^2}+\frac{((8n+1)m_1+3m_2)m_2^2}{12M^4}+\frac{m_1m_2^4}
{9M^6}\right] e^{-m_2^2/M^2}\nonumber  \\
&+g_s^2 \langle \bar{q}_1q_1\rangle^2\bigg [\frac{-2(2n-1)^2-14n+5}
{81M^2}-\frac{m_2^2}{27M^4}\bigg ]e^{-m_2^2/M^2}\nonumber \\
&+4\pi \alpha_s \langle \bar{q}_1q_1\rangle \langle \bar{q}_2q_2 \rangle \frac{4}{9}\left[\frac{2e^{-m_2^2/M^2}}{M^2}+\frac{1}{m_2^2}
(e^{-m_2^2/M^2}-1)\right]\nonumber \\
&+4\pi \alpha_s \langle \bar{q}_1q_1\rangle \langle \bar{q}_2q_2\rangle
\frac{2}{9(m_2^2-m_1^2)}\left(e^{-m_1^2/M^2}-e^{-m_2^2/M^2}  \right) \nonumber \\
&+4\pi \alpha_s \langle \bar{q}_1q_1\rangle \langle \bar{q}_2q_2\rangle\frac{4m_1 m_2}{9(m_2^2-m_1^2)} \left(\frac{e^{-m_1^2/M^2}-1}{m_1^2}-\frac{e^{-m_2^2/M^2}-1}{m_2^2}\right)-\left[q_1 \leftrightarrow q_2, m_1 \leftrightarrow m_2 \right] \bigg \}.
\label{eq:oddmomentsd}
\end{align}
The sum rule for the odd moments of $\phi_{S}^{\sigma}$ up to dimension-six condensates is given by
\begin{align}
&-\frac{1}{3}{m_{a_0 (980)}^2} \bar{f}_S^2 e^{-{m_{a_0 (980)}^2}/M^2}\langle \xi^{2n+1}_{\sigma}\rangle \nonumber \\
&= \frac{3}{4\pi^2}\int^1_0dx(2x-1)^{2n+1}M^4x(x-1)e^{-\frac{m^2_{12}}{M^{2}x(1-x)}}\nonumber \\
& -\frac{3}{4 \pi^2}\int^1_0dx (2x-1)^{2n+1}\left[x(x-1)\left(1+\frac{S_{\sigma}} { M^2}\right) +\frac{m^2_{12}}{M^2}\right] M^4 e^{-{S_{\sigma}/M^2}} \nonumber  \\
&-\langle \alpha_s G^2\rangle \int^1_0dx\frac{(2x-1)^{2n+1}}{24\pi}
\left[ 1-\frac{2m_1m_2}{M^2x(1-x)} \right] e^{-\frac{m^2_{12}}{M^{2}x(1-x)}} \nonumber\\
&+\langle g_s^3fG^3\rangle \frac{m_1m_2}{24\pi^2}\int^1_0dx
\frac{(2x-1)^{2n+1}}{2M^4x(1-x)}e^{-\frac{m^2_{12}}{M^2x(1-x)}}\nonumber\\
&+\bigg \{\langle \bar{q}_1q_1\rangle \frac{-1}{6}\bigg [ 3m_1+\frac{(4n+3)m_1^3}
{M^2}+\frac{m_1^3m_2^2}{M^4}\bigg ] e^{-m_2^2/M^2}\nonumber\\
&+\langle g_s \bar{q}_1 \sigma TGq_1\rangle \bigg [
\frac{(16n+9)m_1+6m_2}{36M^2}+\frac{m_1m_2^2}{9M^4}\bigg ]e^{-m_2^2/M^2}\nonumber \\
&+\frac{g_s^2\langle \bar{q}_1q_1\rangle^2}{81}\bigg [ \frac{-4n+1}{M^2}+\frac{2m_2^2}{M^4}\bigg ]
 e^{-m_2^2/M^2}
-\left[q_1 \leftrightarrow q_2, m_1 \leftrightarrow m_2\right] \bigg \},
\label{eq:oddmomentssigma}
\end{align}
where $M$ is the Borel parameter, $m_{12}^2=m_{1}^{2} x + m_{2}^{2} (1-x)$, $S_s$ and $S_{\sigma}$ are the threshold parameters which are taken to be around the squared mass of the scalar's first excited state.
\end{appendix}


\begin{thebibliography}{100}

\bibitem{Cheng:2005nb}
  H.~Y.~Cheng, C.~K.~Chua and K.~C.~Yang,
  Phys.\ Rev.\ D {\bf 73}, 014017 (2006)
  doi:10.1103/PhysRevD.73.014017
  [hep-ph/0508104].

\bibitem{Sun:2010nv}
  Y.~J.~Sun, Z.~H.~Li and T.~Huang,
  Phys.\ Rev.\ D {\bf 83}, 025024 (2011)
  doi:10.1103/PhysRevD.83.025024
  [arXiv:1011.3901 [hep-ph]].

\bibitem{Mathur:2006bs}
  N.~Mathur {\it et al.},
  Phys.\ Rev.\ D {\bf 76}, 114505 (2007)
  doi:10.1103/PhysRevD.76.114505
  [hep-ph/0607110].

\bibitem{Prelovsek:2010kg}
  S.~Prelovsek, T.~Draper, C.~B.~Lang, M.~Limmer, K.~F.~Liu, N.~Mathur and D.~Mohler,
  Phys.\ Rev.\ D {\bf 82}, 094507 (2010)
  doi:10.1103/PhysRevD.82.094507
  [arXiv:1005.0948 [hep-lat]].

\bibitem{Alexandrou:2012rm}
  C.~Alexandrou, J.~O.~Daldrop, M.~Dalla Brida, M.~Gravina, L.~Scorzato, C.~Urbach and M.~Wagner,
  JHEP {\bf 1304}, 137 (2013)
  doi:10.1007/JHEP04(2013)137
  [arXiv:1212.1418 [hep-lat]].

\bibitem{Cheng:2013fba}
  H.~Y.~Cheng, C.~K.~Chua, K.~C.~Yang and Z.~Q.~Zhang,
  Phys.\ Rev.\ D {\bf 87}, no. 11, 114001 (2013)
  doi:10.1103/PhysRevD.87.114001
  [arXiv:1303.4403 [hep-ph]].

\bibitem{Cheng:2003sm}
  H.~Y.~Cheng, C.~K.~Chua and C.~W.~Hwang,
  Phys.\ Rev.\ D {\bf 69}, 074025 (2004)
  doi:10.1103/PhysRevD.69.074025
  [hep-ph/0310359].
%
\bibitem{Wang:2008da}
  Y.~M.~Wang, M.~J.~Aslam and C.~D.~Lu,
  Phys.\ Rev.\ D {\bf 78}, 014006 (2008)
  doi:10.1103/PhysRevD.78.014006
  [arXiv:0804.2204 [hep-ph]].

\bibitem{Wirbel:1985ji}
  M.~Wirbel, B.~Stech and M.~Bauer,
  Z.\ Phys.\ C {\bf 29}, 637 (1985).
  doi:10.1007/BF01560299

\bibitem{Cheung:1995ub}
  C.~Y.~Cheung, W.~M.~Zhang and G.~L.~Lin,
  Phys.\ Rev.\ D {\bf 52}, 2915 (1995)
  doi:10.1103/PhysRevD.52.2915
  [hep-ph/9505232].

\bibitem{Zhang:1994hg}
  W.~M.~Zhang, G.~L.~Lin and C.~Y.~Cheung,
  Int.\ J.\ Mod.\ Phys.\ A {\bf 11}, 3297 (1996)
  doi:10.1142/S0217751X96001577
  [hep-ph/9412394].

\bibitem{Choi:1999nu}
  H.~M.~Choi and C.~R.~Ji,
  Phys.\ Lett.\ B {\bf 460}, 461 (1999)
  doi:10.1016/S0370-2693(99)00817-5
  [hep-ph/9903496].

\bibitem{Shifman:1978by}
  M.~A.~Shifman, A.~I.~Vainshtein and V.~I.~Zakharov,
  Nucl.\ Phys.\ B {\bf 147}, 448 (1979).
  doi:10.1016/0550-3213(79)90023-3
%
\bibitem{Novikov:1981xi}
  V.~A.~Novikov, M.~A.~Shifman, A.~I.~Vainshtein and V.~I.~Zakharov,
  Nucl.\ Phys.\ B {\bf 191}, 301 (1981).
  doi:10.1016/0550-3213(81)90303-5

\bibitem{Balitsky:1989ry}
  I.~I.~Balitsky, V.~M.~Braun and A.~V.~Kolesnichenko,
  Nucl.\ Phys.\ B {\bf 312}, 509 (1989).
  doi:10.1016/0550-3213(89)90570-1

\bibitem{Braun:1988qv}
  V.~M.~Braun and I.~E.~Filyanov,
  Z.\ Phys.\ C {\bf 44}, 157 (1989)
  [Sov.\ J.\ Nucl.\ Phys.\  {\bf 50}, 511 (1989)]
  [Yad.\ Fiz.\  {\bf 50}, 818 (1989)].
  doi:10.1007/BF01548594

\bibitem{Chernyak:1990ag}
  V.~L.~Chernyak and I.~R.~Zhitnitsky,
  Nucl.\ Phys.\ B {\bf 345}, 137 (1990).
  doi:10.1016/0550-3213(90)90612-H

\bibitem{Keum:2000ph}
  Y.~Y.~Keum, H.~n.~Li and A.~I.~Sanda,
  Phys.\ Lett.\ B {\bf 504}, 6 (2001)
  doi:10.1016/S0370-2693(01)00247-7
  [hep-ph/0004004].

\bibitem{Keum:2000wi}
  Y.~Y.~Keum, H.~N.~Li and A.~I.~Sanda,
  Phys.\ Rev.\ D {\bf 63}, 054008 (2001)
  doi:10.1103/PhysRevD.63.054008
  [hep-ph/0004173].

\bibitem{Lu:2000em}
  C.~D.~Lu, K.~Ukai and M.~Z.~Yang,
  Phys.\ Rev.\ D {\bf 63}, 074009 (2001)
  doi:10.1103/PhysRevD.63.074009
  [hep-ph/0004213].
%
\bibitem{Lu:2006fr}
  C.~D.~Lu, Y.~M.~Wang and H.~Zou,
  Phys.\ Rev.\ D {\bf 75}, 056001 (2007)
  doi:10.1103/PhysRevD.75.056001
  [hep-ph/0612210].

\bibitem{Han:2013zg}
  H.~Y.~Han, X.~G.~Wu, H.~B.~Fu, Q.~L.~Zhang and T.~Zhong,
  Eur.\ Phys.\ J.\ A {\bf 49}, 78 (2013)
  doi:10.1140/epja/i2013-13078-7
  [arXiv:1301.3978 [hep-ph]].

\bibitem{Onyisi:2013bjt}
  P.~U.~E.~Onyisi {\it et al.} [CLEO Collaboration],
  Phys.\ Rev.\ D {\bf 88}, no. 3, 032009 (2013)
  doi:10.1103/PhysRevD.88.032009
  [arXiv:1306.5363 [hep-ex]].

\bibitem{Bonvicini:2013vxi}
  G.~Bonvicini {\it et al.} [CLEO Collaboration],
  Phys.\ Rev.\ D {\bf 89}, no. 7, 072002 (2014)
  Erratum: [Phys.\ Rev.\ D {\bf 91}, no. 1, 019903 (2015)]
  doi:10.1103/PhysRevD.89.072002, 10.1103/PhysRevD.91.019903
  [arXiv:1312.6775 [hep-ex]].

\bibitem{Asner:2008nq}
  D.~M.~Asner {\it et al.},
  Int.\ J.\ Mod.\ Phys.\ A {\bf 24}, S1 (2009)
  [arXiv:0809.1869 [hep-ex]].

\bibitem{Cheng:2007uj}
  X.~D.~Cheng, K.~L.~He, H.~B.~Li, Y.~F.~Wang and M.~Z.~Yang,
  Phys.\ Rev.\ D {\bf 75}, 094019 (2007)
  doi:10.1103/PhysRevD.75.094019
  [arXiv:0704.0120 [hep-ex]].

\bibitem{Ablikim:2015wel}
  M.~Ablikim {\it et al.} [BESIII Collaboration],
  Phys.\ Rev.\ Lett.\  {\bf 116}, no. 8, 082001 (2016)
  doi:10.1103/PhysRevLett.116.082001
  [arXiv:1512.06998 [hep-ex]].

\bibitem{Wang:2009azc}
  W.~Wang and C.~D.~Lu,
  Phys.\ Rev.\ D {\bf 82}, 034016 (2010)
  doi:10.1103/PhysRevD.82.034016
  [arXiv:0910.0613 [hep-ph]].

\bibitem{Chernyak:1983ej}
  V.~L.~Chernyak and A.~R.~Zhitnitsky,
  Phys.\ Rept.\  {\bf 112}, 173 (1984).
  doi:10.1016/0370-1573(84)90126-1

\bibitem{Braun:2003rp}
  V.~M.~Braun, G.~P.~Korchemsky and D.~Mueller,
  Prog.\ Part.\ Nucl.\ Phys.\  {\bf 51}, 311 (2003)
  doi:10.1016/S0146-6410(03)90004-4
  [hep-ph/0306057].

\bibitem{Balitsky:1987bk}
  I.~I.~Balitsky and V.~M.~Braun,
  Nucl.\ Phys.\ B {\bf 311}, 541 (1989).
  doi:10.1016/0550-3213(89)90168-5

\bibitem{Khodjamirian:1998ji}
  A.~Khodjamirian and R.~Ruckl,
  Adv.\ Ser.\ Direct.\ High Energy Phys.\  {\bf 15}, 345 (1998)
  doi:10.1142/9789812812667$_{-}$0005
  [hep-ph/9801443].

\bibitem{Diehl:1998kh}
  M.~Diehl, T.~Feldmann, R.~Jakob and P.~Kroll,
  Eur.\ Phys.\ J.\ C {\bf 8}, 409 (1999)
  doi:10.1007/s100529901100
  [hep-ph/9811253].

\bibitem{Colangelo:2010bg}
  P.~Colangelo, F.~De Fazio and W.~Wang,
  Phys.\ Rev.\ D {\bf 81}, 074001 (2010)
  doi:10.1103/PhysRevD.81.074001
  [arXiv:1002.2880 [hep-ph]].

\bibitem{Hambrock:2015aor}
  C.~Hambrock and A.~Khodjamirian,
  Nucl.\ Phys.\ B {\bf 905}, 373 (2016)
  doi:10.1016/j.nuclphysb.2016.02.035
  [arXiv:1511.02509 [hep-ph]].

\bibitem{Narison:2010wb}
  S.~Narison,
  Nucl.\ Phys.\ Proc.\ Suppl.\  {\bf 207-208}, 315 (2010)
  doi:10.1016/j.nuclphysbps.2010.10.078
  [arXiv:1010.1959 [hep-ph]].

\bibitem{Narison:2011rn}
  S.~Narison,
  Phys.\ Lett.\ B {\bf 707}, 259 (2012)
  doi:10.1016/j.physletb.2011.12.047
  [arXiv:1105.5070 [hep-ph]].

\bibitem{Olive:2016xmw}
  C.~Patrignani {\it et al.} [Particle Data Group],
  Chin.\ Phys.\ C {\bf 40}, no. 10, 100001 (2016).
  doi:10.1088/1674-1137/40/10/100001

\bibitem{Navarra:2000ji}
  F.~S.~Navarra, M.~Nielsen, M.~E.~Bracco, M.~Chiapparini and C.~L.~Schat,
  Phys.\ Lett.\ B {\bf 489}, 319 (2000)
  doi:10.1016/S0370-2693(00)00967-9
  [hep-ph/0005026].

\bibitem{Dosch:2002rh}
  H.~G.~Dosch, E.~M.~Ferreira, F.~S.~Navarra and M.~Nielsen,
  Phys.\ Rev.\ D {\bf 65}, 114002 (2002)
  doi:10.1103/PhysRevD.65.114002
  [hep-ph/0203225].

\bibitem{Matheus:2002nq}
  R.~D.~Matheus, F.~S.~Navarra, M.~Nielsen and R.~Rodrigues da Silva,
  Phys.\ Lett.\ B {\bf 541}, 265 (2002)
  doi:10.1016/S0370-2693(02)02259-1
  [hep-ph/0206198].

\bibitem{Bracco:2004rx}
  M.~E.~Bracco, M.~Chiapparini, F.~S.~Navarra and M.~Nielsen,
  Phys.\ Lett.\ B {\bf 605}, 326 (2005)
  doi:10.1016/j.physletb.2004.11.024
  [hep-ph/0410071].

\bibitem{Lucha:2009uy}
  W.~Lucha, D.~Melikhov and S.~Simula,
  Phys.\ Rev.\ D {\bf 79}, 096011 (2009)
  doi:10.1103/PhysRevD.79.096011
  [arXiv:0902.4202 [hep-ph]].

\end{thebibliography}
\end{document}